\documentclass{aa}

\usepackage{graphicx}
\usepackage{xcolor}

\usepackage{txfonts}
\usepackage{hyperref}
\usepackage[version=4]{mhchem}

\begin{document}

   \title{The Asymmetric Bipolar [Fe II] Jet and \ce{H2} Outflow of TMC1A Resolved with JWST's NIRSpec IFU}

   \subtitle{}

   \author{K. D. Assani \inst{1}
          \and D. Harsono\inst{2}
          \and J. P. Ramsey \inst{1}
          \and Z.-Y. Li \inst{1}
          \and P. Bjerkeli \inst{3}
          \and K. M. Pontoppidan \inst{4}
          \and Ł. Tychoniec \inst{5}
          \and H. Calcutt \inst{3}
          \and L. E. Kristensen \inst{7}
          \and J. K. J\o{}rgensen \inst{7}
          \and A. Plunkett \inst{8}
          \and M. L. van Gelder \inst{5}
          \and L. Francis \inst{5}
          }

   \institute{Department of Astronomy, University of Virginia,
              Charlottesville, VA 22903, USA
               \and  
             Institute of Astronomy, Department of Physics, National Tsing Hua University, Hsinchu, Taiwan
             \and
             Chalmers University of Technology, Department of Space, Earth and Environment, SE-412 96 Gothenburg, Sweden 
             \and 
             Jet Propulsion Laboratory, California Institute of Technology, 4800 Oak Grove Drive, Pasadena, CA 91109, USA 
             \and 
             Space Telescope Science Institute, 3700 San Martin Drive, Baltimore, MD 21218, USA
             \and 
             Leiden Observatory, Leiden University, PO Box 9513, 2300RA, Leiden, The Netherlands
             \and 
             Niels Bohr Institute, University of Copenhagen, \O{}ster Voldgade 5–7, DK 1350 Copenhagen K., Denmark
             \and 
             National Radio Astronomy Observatory, 520 Edgemont Road, Charlottesville, VA 22903, USA
             }

   \date{Received 2024, Accepted 2024}

 
  \abstract
   {Protostellar outflows exhibit large variations in their structure depending on the observed gas emission. To understand the origin of the observed variations, it is important to analyze differences in the observed morphology and kinematics of the different tracers. The {\it James Webb Space Telescope} ({\it JWST}) allows us to study the physical structure of the protostellar outflow through well-known near-infrared shock tracers in a manner unrivaled by other existing ground-based and space-based telescopes at these wavelengths.}
   {This study analyzes the atomic jet and molecular outflow in the Class I protostar, TMC1A, utilizing spatially resolved [\ion{Fe}{II}] and \ce{H2} lines to characterize morphology and identify previously undetected spatial features, and compare them to existing observations of TMC1A and its outflows observed at other wavelengths.}
   {We identify a large number of [\ion{Fe}{II}] and \ce{H2} lines within the G140H, G235H, and G395H gratings of the NIRSpec IFU observations. We analyze their morphology and position-velocity (PV) diagrams. From the observed [\ion{Fe}{II}] line ratios, the extinction toward the jet is estimated.}
   {We have detected the bipolar Fe jet by revealing, for the first time, the presence of a red-shifted atomic jet. Similarly, the red-shifted component of the \ce{H2} slower wide-angle outflow is observed. Both [\ion{Fe}{II}] and \ce{H2} red-shifted emission exhibit significantly lower flux densities compared to their blue-shifted counterparts. Additionally, we report the detection of a collimated high-velocity ($\sim 100$ km s$^{-1}$), blue-shifted \ce{H2} outflow, suggesting the presence of a molecular jet in addition to the well-known wider angle low-velocity structure. The [\ion{Fe}{II}] and \ce{H2} jets show multiple intensity peaks along the jet axis, which may be associated with ongoing or recent outburst events. In addition to the variation in their intensities, the \ce{H2} wide-angle outflow exhibits a "ring"-like structure. The blue-shifted \ce{H2} outflow also shows a left-right brightness asymmetry likely due to interactions with the surrounding ambient medium and molecular outflows. Using the [\ion{Fe}{II}] line ratios, the extinction along the atomic jet is estimated to be between $A_{\rm V}$ = 10--30 on the blue-shifted side, with a trend of decreasing extinction with distance from the protostar. A similar $A_{\rm V}$ is found for the red-shifted side, supporting the argument for an intrinsic red-blue outflow lobe asymmetry rather than environmental effects such as extinction. 
   This intrinsic difference revealed by the unprecedented sensitivity of {\it JWST}, suggests that younger outflows already exhibit the red-blue side asymmetry more commonly observed towards jets associated with Class II disks.
   }
   {}


   \maketitle
%
\section{Introduction} \label{introduction}

Protostellar outflows are an integral part of star formation, transporting material, momentum, and energy from the protostar and accretion disk, and contributing to the redistribution of mass and angular momentum throughout the protostellar system \citep[e.g][]{franketal14_ppvi, myers2023_outflows, pascucci2023_outflow}. There is a great deal of diversity in the structure, strength, and even presence of atomic and molecular emission lines in protostellar outflows, alluding to a possible diversity in the launching mechanisms \citep[e.g.][]{arce2007,Seale_morphology_diff_bipolar_2008,Bally_protostellar_outflows_2016,Tychoniec_serpens_jets_2019}. One way to improve our understanding of protostellar outflows is by analyzing spatially resolved near-IR atomic and molecular emission along the protostellar outflow using state-of-the-art instruments such as the {\it James Webb Space Telescope} ({\it JWST}).

Most protostellar outflows are observed to be bipolar \citep[e.g.][]{Shu_star_formation_bipolar_1987, Frank_bipolar_outflows_1999, bally_PPV_2007, Seale_morphology_diff_bipolar_2008,Bally_protostellar_outflows_2016}, but are often asymmetric in their brightness and morphology  \citep[e.g.][]{Mundt_optical_jets_asym_1990,Ray_HST_jets_asym_1996,Ray_PPV_outflow_asym_2007,white_asymmetry_2014_dgtau,Bally_protostellar_outflows_2016}. It is still not understood whether observed differences in the sides/lobes of bipolar outflows are due to intrinsic differences in the strengths (which we correlate with brightness) of the outflows, or due to extrinsic factors such as a non-uniform ambient medium. There have been several attempts to explain these asymmetries around more evolved (Class II) sources. For example, in the case of FS Tauri B, \cite{Liu_2012_asymmetry} suggested that the difference in mass-loss rate found in the bipolar outflow lobes is a result of the system maintaining linear momentum balance, which results in the lobe with less mass having a higher velocity. In the case of DG Tauri B, \cite{podio_2011_asymmetry_dgtau} suggests the bipolar asymmetry is due to the interaction of the bipolar outflow with an asymmetric ambient medium, while \cite{white_asymmetry_2014_dgtau} proposed that the outflow might be intrinsically symmetric with the observed asymmetry caused by environmental effects such as obscuration by an asymmetric ambient medium. Furthermore, asymmetries between the two bipolar lobes have been observed in a number of Class II sources, such as AW Aurigae \citep[e.g.][]{wowitas2002} and TH 28 \citep[e.g.][]{melnikov2023}. With {\it JWST}, we can establish whether such asymmetry is observed toward the more deeply embedded protostars.

Protostellar outflows that are collimated and travel at velocities greater than $\sim$100 km s$^{-1}$ are called protostellar jets \citep[e.g.][]{Bally_protostellar_outflows_2016,ray2023outflows}. While jets are observed at all evolutionary stages in protostellar systems with active accretion, their overall strengths and the presence of collimated, fast, molecular components tend to decline with age \citep[e.g.][]{bontemps_outflow_evolution_1996,reipurthbally01,cabrit_molecular_atomic_jets_2011, Bally_protostellar_outflows_2016,podio_jet_ubiquitous_2021}. Molecular jets are typically associated with younger protostellar sources such as the deeply embedded Class 0 protostar, HH 211, where a fast ($\sim$100 km s$^{-1}$) \ce{H2} jet was found \citep{ray2023outflows} in addition to a knotty and bipolar jet in SiO \citep{Lee_Li_HH211_SiO_2018}. However, a deeply embedded atomic jet in [\ion{Fe}{II}], [\ion{S}{I}], [\ion{S}{II}] has also been detected in HH 211 with Spitzer \citep{Dionatos_H211_atomic_jet_2010}. An embedded jet is also detected in the Class 0 protostar, L1448-C, but the atomic lines are less bright and carry less overall mass flux than their molecular counterparts \citep{Dionatos_atomic_molecular_l1448_2009, Nisini_Class_0_jets_2015}. In more evolved Class I sources, atomic jets tend to be observed carrying more mass flux at higher velocities and at larger scales than the molecular jets \citep{davis_classI_FEL_2003,nisini_hh1_jet_2005, podio2006recipes,Sperling_atomic_evolution_2021}. Finally, in Class II sources, the jet as a whole becomes fainter, and only the atomic component is found to still be well-collimated \citep{Hartigan_accretion_mass_loss_1995,Hirth_atomicjets_Ttauri_1997}, while the molecular component is slower and no longer well-collimated \citep{beck2008spatially, agra_amboage_wideangleh2_dgtau_2014}.

These observed trends may point to the survival of molecules in the jet such as \ce{H2} in deeply embedded Class 0 protostars due to shielding by dust \citep[e.g.,][]{cabrit_molecular_atomic_jets_2011,panoglouetal2012,Tabone_H2_formation_shielding_2020}. As the protostellar system evolves, the mass accretion rate onto the protostar decreases, the protostellar envelope dissipates, and ultra-violet (UV) photons from the star may destroy most of the molecules. Alternatively, it might be possible that when infall is highest near the beginning of the star formation process, the launched outflows may be denser and slower, resulting in weaker shocks where molecules can survive or reform more easily.

Along with the atomic and molecular jets, large-scale spatial brightness variations along protostellar jets, typically referred to as ``knots'', have been observed. Along with shocks, these knots have historically been designated as Herbig-Haro (HH) objects \citep[e.g.][]{herbig51,Haro1952,reipurthbally01}. Knots are usually explained as either high-velocity jet material colliding with the slower surrounding molecular gas or older, slower outflow material, creating shock fronts \citep[e.g.][]{mundt_fried_1983_4jets}. A well-studied example is the HH 111 complex which shows a high-velocity [\ion{S}{II}] and H$\alpha$ jet \citep[e.g.][]{reipurth_1997_HST_H111} and high-velocity ($\sim$400-500 km s$^{-1}$) molecular CO "bullets" \citep{Cernicharo_Reipurth_1996_bullets} in addition to \ce{H2} knots situated in between the other components and the bow shock structure \citep{gredel1994near}. The CO "bullets" and \ce{H2} knots are thought to be the dense material trapped between the radiative shocks which have cooled sufficiently to form molecules \citep{reipurth_50years_HH,Melnick_O2_formation_2015,Godard_molecular_shock_model_2019}.

{\it JWST} provides unprecedented access to near- and mid-infrared wavelengths that are home to many atomic and molecular lines, including forbidden [\ion{Fe}{II}] and rotational \ce{H2} emission lines known to be shock tracers \citep[e.g.][]{giannini2013diagnostic, Reipurthetal2019, Fedrianietal2019, harsono2023_tmc1a, federman2023investigating, beuther_joys_2023,Lukasz_what_traces_what_2021,Delabrosse_JWST_DGtauB_2024,Nisini_JWST_HH46_47_2024}. Specifically, the NIRSpec Integral Field Unit (IFU) \citep{jakobsen2022near, boker2022near} on JWST covers a $\sim$3x3" field of view (FOV) with a $0\farcs{1}$ spatial resolution, providing a high-quality spectrum at each spaxel. The NIRSpec IFU has already proven a valuable instrument for revealing previously undetected jets in Class 0/1 sources such as IRAS 16253-2429 \citep{narang2023investigating} and TMC1A \citep{harsono2023_tmc1a}. Previous JWST-NIRSpec studies have primarily focused on longer wavelength [Fe II] lines such as 4.11 and 4.89 $\mu m$ with the G395M/H gratings, and the 5.34 and 17.89 $\mu m$ lines with the MIRI Medium Resolution Spectrometer (MRS) \citep{rieke2015mid, wright2015mid, yangetal2022}. However, there are numerous [\ion{Fe}{II}] lines in 1--2 $\mu$m region that can be used to trace shocks and excitation conditions \citep{giannini_2008_nearIR_jets, giannini2013diagnostic, giannini2015empirical, koo_2016_shock_models}. 

The Class I low-mass protostar TMC1A, observed with NIRSpec IFU from 1-5 $\mu$m inclusive, presents a great opportunity to investigate the aforementioned emission lines that trace shocks of both atomic and molecular outflows. Additional data is also available as the molecular outflow of TMC1A has been observed by the Atacama Large Millimeter/sub-millimeter Array, in particular, the outflow traced by the rotational transitions of $^{12}$CO and its isotopologues \citep[e.g.][]{chandleretal1996_tmc1a, aso15, bjerkeli2016, aso2021}. Comparing the different outflow components observed in a single source (such as TMC1A) provides an opportunity to discover how these components differ in terms of kinematics and location.

In this study, we conduct a thorough search for [\ion{Fe}{II}] and \ce{H2} emission lines using {\it JWST}'s NIRSpec IFU, and explore the spatial and kinematic characteristics (e.g.\ morphology) of the atomic iron jet and molecular \ce{H2} outflow associated with the Class I protostar, TMC1A (IRAS 04365+2535, with a distance of 140 pc, \citealt{galli2019structure}). In Section \ref{sec:observations}, we describe the observations and data reduction used in our analysis. In Section \ref{sec:iron_h2_detections}, we discuss the characteristics of the detected [\ion{Fe}{II}] and \ce{H2} lines. In section \ref{sec:atomicjet} and \ref{sec:molecular_jet_and_outflows}, we examine the morphology of the [\ion{Fe}{II}] and \ce{H2} lines. In section \ref{sec:extinction}, we estimate the line-of-sight extinction toward the jet using the [\ion{Fe}{II}] lines. In Section \ref{sec:discussion}, we discuss the implications of our results in the context of protostellar outflows. Finally, in Section \ref{sec:conclusion}, we summarize our results and list our conclusions on the morphology and asymmetries observed in the atomic and molecular outflows.

\section{Observations and Data Reduction}
\label{sec:observations}
\begin{figure*}
\centering
\includegraphics[width=\textwidth,trim=0cm 7cm 0cm 7cm, clip]{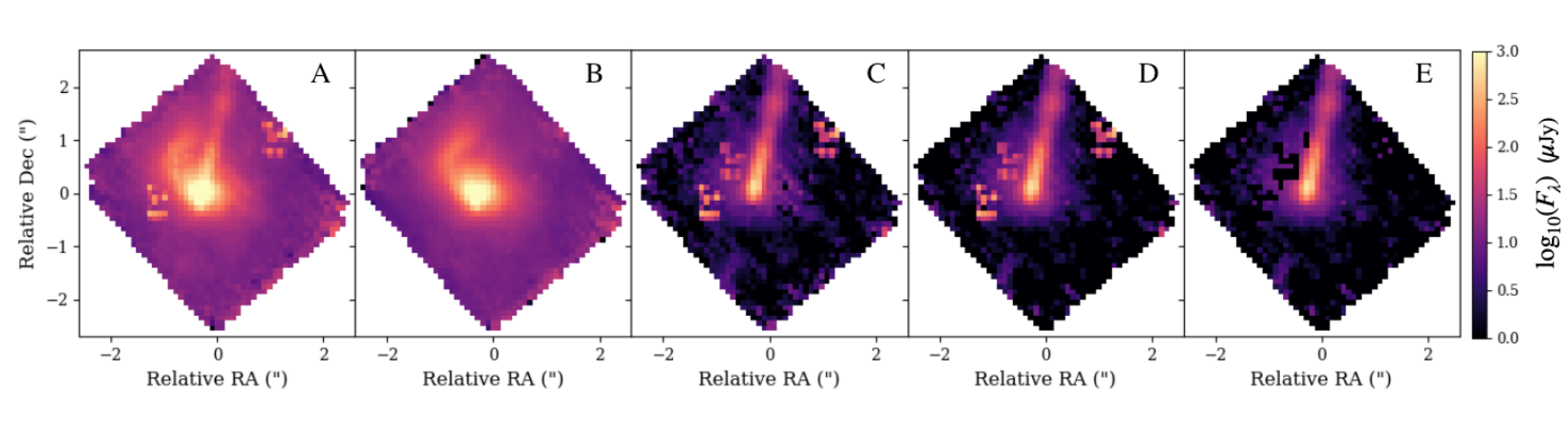}
\caption{Illustration of the data reduction process for the 1.644 $\mu$m [\ion{Fe}{II}] line. (A) Integrated flux density maps (moment 0) of the line after the pipeline. (B) The result of the continuum fit. (C) The continuum-subtracted emission map. (D) The continuum-subtracted map with an SNR cut ($F_{\lambda} \geq 3$). (E) The zeroth moment map after removal of bad spaxels. } 
\label{fig:figsreduction}
\end{figure*}

{\it JWST}-NIRSpec observations of TMC1A were first presented in \cite{harsono2023_tmc1a}  (PID: 2104, PI: Harsono), and we refer the reader to that paper for the observational details. Here, we focus our analysis on the 1--5 $\mu$m [\ion{Fe}{II}] and \ce{H2} emission lines. We use the \href{https://linelist.pa.uky.edu/newpage/}{Atomic Line List version:3.00b5 database} \citep{van_Hoof_atomic_lines_2018} and the \href{https://physics.nist.gov/PhysRefData/ASD/lines_form.html}{NIST Atomic Spectra Database} \citep{NIST_ASD_2023} to find the physical properties of each [\ion{Fe}{II}] transition (rest wavelengths, Einstein $A$ coefficients, energy levels), and the \href{https://www.gemini.edu/observing/resources/near-ir-resources/spectroscopy/important-h2-lines}{Gemini \ce{H2} line list} to obtain the corresponding information for the \ce{H2} lines. The velocity of the line emission in each spaxel is determined relative to the vacuum rest wavelengths and calculated using the formula $v_i = \frac{\lambda_i - \lambda_0}{\lambda_0}c$, where $\lambda_0$ represents the rest wavelength and $c$ is the speed of light. We adopt the rest wavelengths in the Atomic Line List Database for the [\ion{Fe}{II}] lines, although we note that differences in the rest wavelength can lead to different velocities.

In \citet{harsono2023_tmc1a}, the Stage 3 pipelined products were generated with \textsc{JWST} pipeline v1.9.6 \citep[][]{bushouse_2023_1_9_6}. In this study, we use v1.11.4 pipeline \citep[][]{bushouse_2023_1_11_4} and have re-calibrated the raw \textit{JWST} data using the updated calibration files \textsc{jwst\_1123.pmap} \citep{jwstcrds}.  The new pipeline has improved upon the \textit{snowball} artifact that is caused by the cosmic rays and its flagging routines for outlier detections.  Furthermore, a large improvement in flux calibration was seen between V1.9.6 and v1.11.4.  Following \citet{harsono2023_tmc1a} and \citet{Sturm_ERS_2023}, the \href{https://jwst-pipeline.readthedocs.io/en/latest/jwst/outlier_detection/index.html}{outlier detection} step was skipped during the Stage 3 pipeline 
as it tends to flag out strong lines and the saturated central pixel extensively. After inspecting the final spectral cubes, we manually mask the bad spaxels after performing continuum subtraction and applying signal-to-noise constraints.

In Figure \ref{fig:figsreduction}, we show a series of integrated emission (moment 0) maps, centered on the [\ion{Fe}{II}] line at 1.644 $\mu$m to illustrate each step of the data reduction process (above and beyond the \textit{JWST} pipeline). First, we fit the continuum independently in each spaxel within a spectral window near the rest wavelength of each emission line of interest using a linear polynomial fit (Fig.\ \ref{fig:figsreduction}B.). We then subtract this fit from the data in each spaxel to obtain a continuum-subtracted data cube (Fig.\ \ref{fig:figsreduction}C.). Following this, we estimate the noise in each spaxel across the spectral cube by adding together the 1-sigma flux error obtained from the pipeline with a calibration flux error of 5\% in each spaxel (see \href{https://jwst-docs.stsci.edu/jwst-data-calibration-considerations/jwst-calibration-uncertainties}{JWST User Documentation}) and perform a signal-to-noise ratio (SNR) cut to select only data above a threshold (SNR$\geq$3), and mask values below the threshold by setting them to a small value (10$^{-32}$) (Fig, \ref{fig:figsreduction}D.). Bad spaxels \citep{boker2023orbit} appear as delta functions with very large intensities in the spectral dimension. We manually inspect the data to flag these bad spaxels (Fig.\ \ref{fig:figsreduction}E). As seen in the fourth column (D) in Figure \ref{fig:figsreduction}, the location of bad spaxels can differ depending on the wavelength and the location of the spatial mask has to be accommodated accordingly.

We explored several methods for performing the continuum subtraction, e.g., median filters, linear fitting, and taking the average median value of the continuum away from the line. We adopt the linear fitting method because the median filter is sensitive to the size of the kernel and dependent on the wavelength, while the averaging of the median values away from the line does not account for the general increasing trend of the continuum with wavelength, which can be significant from one side of the line to the other. 
Differentiating between emission lines and the continuum poses a difficult challenge near the protostar as the continuum is much brighter thus decreasing the line-to-continuum ratio (i.e.\ $< 50-100$ AU from the protostar in our observations). The issue of contamination due to scattered light is also less significant in the regions far from the protostar since the continuum intensity is much lower. As a result, much of our analysis avoids focusing on the region near the protostar.

\section{Results: The atomic and molecular outflows}
\begin{figure*}
\centering
\includegraphics[width=\textwidth,trim=2cm 4cm 2cm 3cm, clip]{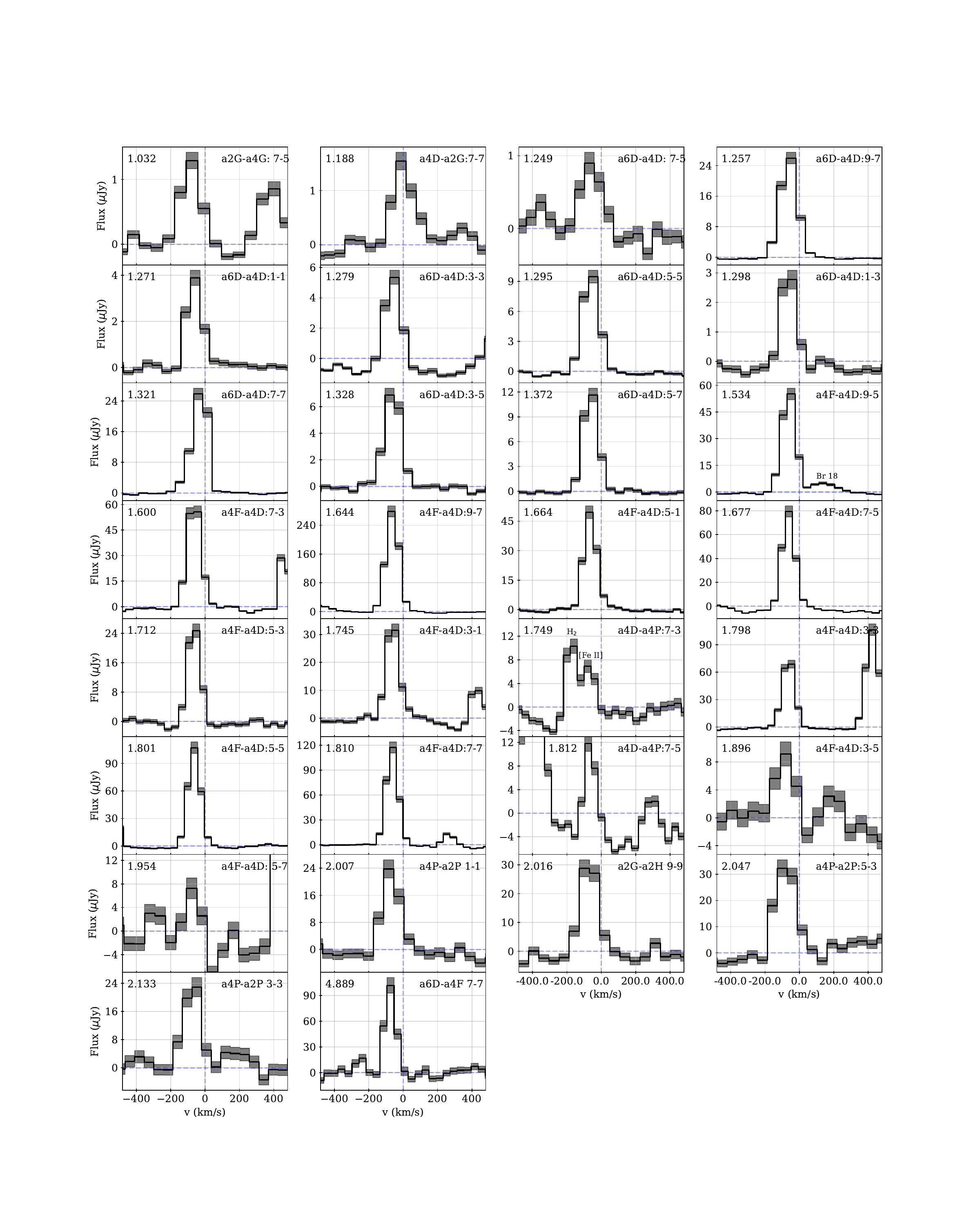}
\caption{Detected [\ion{Fe}{II}] emission lines in order of increasing wavelength. Each panel shows a spectrum taken in the northern part of the outflow along the jet axis ($\sim$0.5 arcseconds away from the protostar).  The location of the spaxel is shown as an "x" in the panel of the 1.644 $\mu m$ line of Fig. \ref{fig:full_iron2lines_moment0}. Meanwhile, the 4.889 $\mu$m spectrum is extracted from a location that is marked by "x" in the panel for the 4.889 $\mu m$ in Fig.\ \ref{fig:full_iron2lines_moment0} to avoid the contamination by the CO fundamental ro-vibrational line. Velocities are calculated independently for each line with respect to the rest wavelength (see Table \ref{tab:combined_detections}). The wavelength in microns is given in the upper left of each panel, and the transition terms are given in the upper right and summarized in Table \ref{tab:combined_detections}. The 1.749 $\mu$m line is blended with the \ce{H2} 1-0 S(7) line and each are labeled in the inset. Similarly the Br 18 hydrogen recombination line is denoted in the 1.534 $\mu$m line.} The 1$\sigma$ uncertainty including a 5\% calibration uncertainty is shown by the shaded region. 
\label{fig:full_iron2lines_spectra}
\end{figure*}

In this section, we describe our detections and analysis of the [\ion{Fe}{II}] and \ce{H2} lines. For each of the strong lines, we analyze their morphology by comparing their flux density maps, spectral line profiles, and position-velocity (PV) diagrams. Finally, we use these results to estimate the extinction along the jet. 

\begin{figure*}
\centering
\includegraphics[width=\textwidth,trim=0cm 0cm 0cm 0cm, clip]{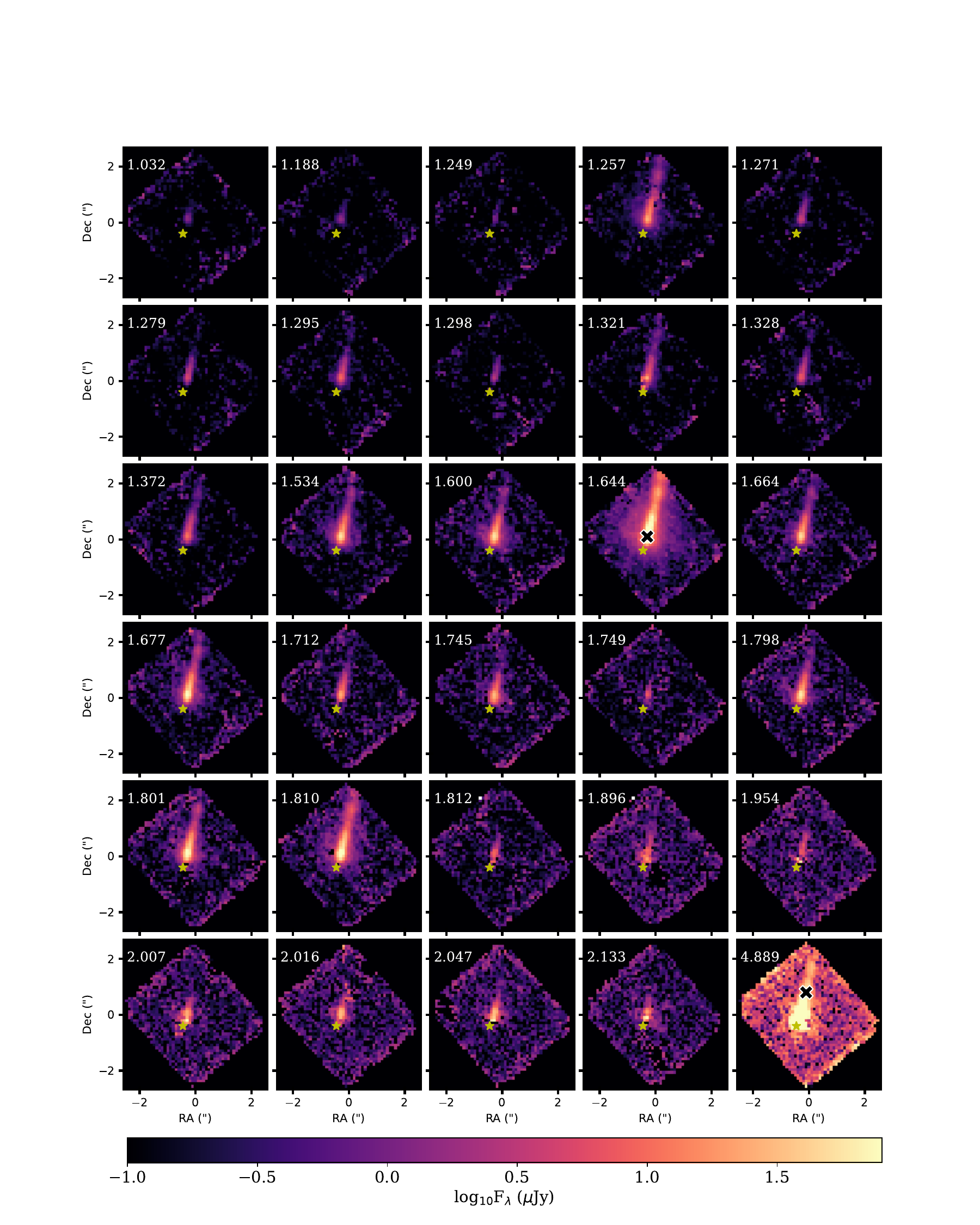}
\caption{Each panel shows the peak flux density map of the detected [\ion{Fe}{II}] lines with an SNR $\geq$ 3 and bad pixels masked out. The black "x" with white borders in the panel for the 1.644 $\mu m$ line denotes the spaxel used to extract the 1D spectra in Figure \ref{fig:full_iron2lines_spectra}. A different spaxel is used to extract the 4.89 $\mu m$ line profile since it is affected by contamination closer to the protostar. The red-shifted or southern part of the outflow discussed in later sections (see Sect.\ \ref{sec:atomicjet}) will not appear in these maps since we plot the flux density maps at the velocity corresponding to the peak line flux for a spaxel located in the blue-shifted side of the jet. The rest wavelength in microns of each line is over-plotted in the upper left of each panel. } 
\label{fig:full_iron2lines_moment0}
\end{figure*} 

\begin{figure*}
\centering
\includegraphics[width=0.935\textwidth,trim=2.2cm 4cm 2.2cm 5cm, clip]{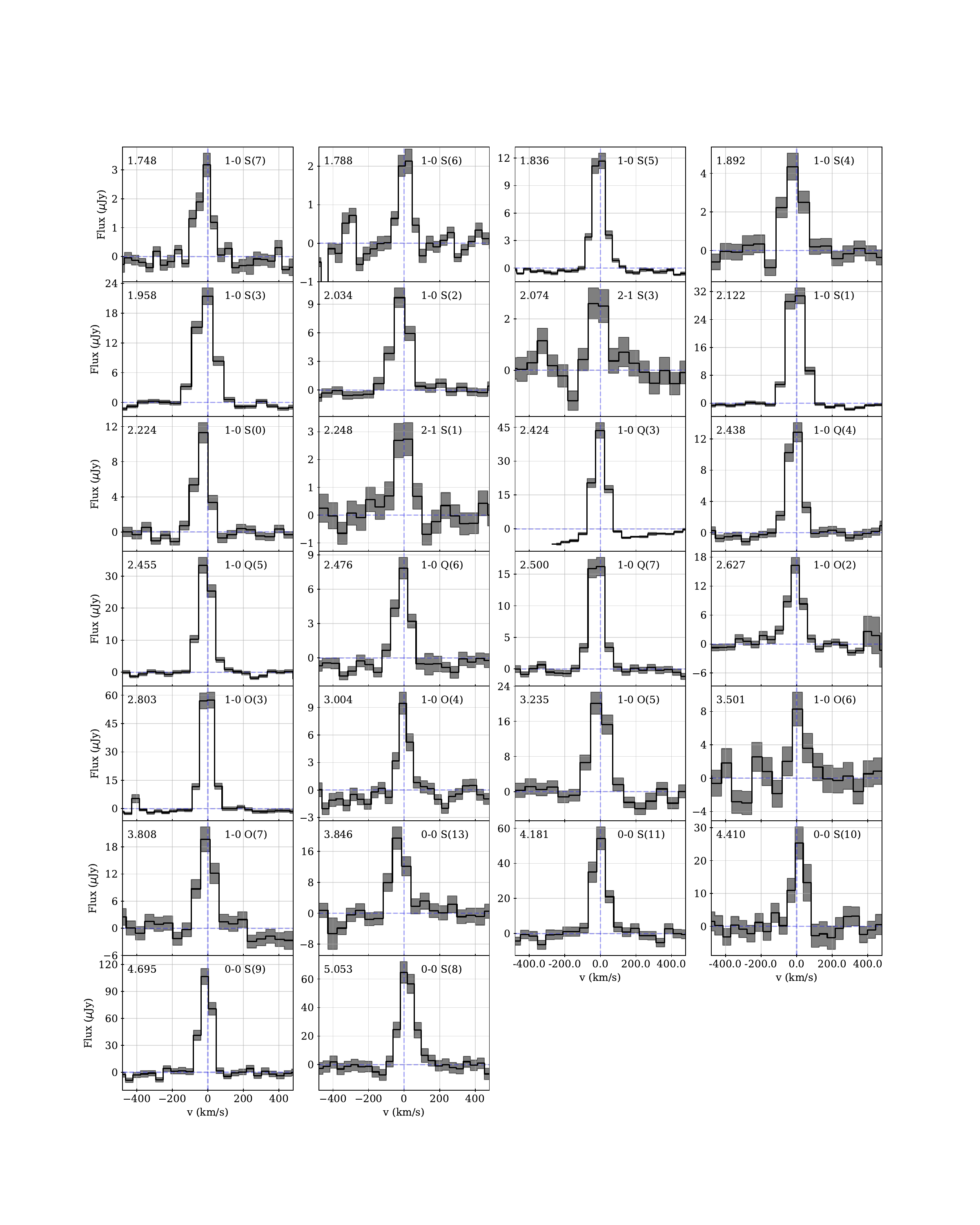}
\caption{Line profiles of the detected \ce{H2} emission lines in order of increasing wavelength. Each spectrum is taken in the northern part of the outflow marked by the black "x" in the 2.122 $\mu$m panel of Fig.\ \ref{fig:full_H2lines_moment0}. This location is chosen because this region is bright in \ce{H2},  while the contamination from scattered light that is worse at longer wavelengths is minimized, easing the detection of weaker line emission. } 
\label{fig:full_H2lines_spectra}
\end{figure*}

\begin{figure*}
\centering
\includegraphics[width=\textwidth,trim=1cm 4cm 1cm 4cm, clip]{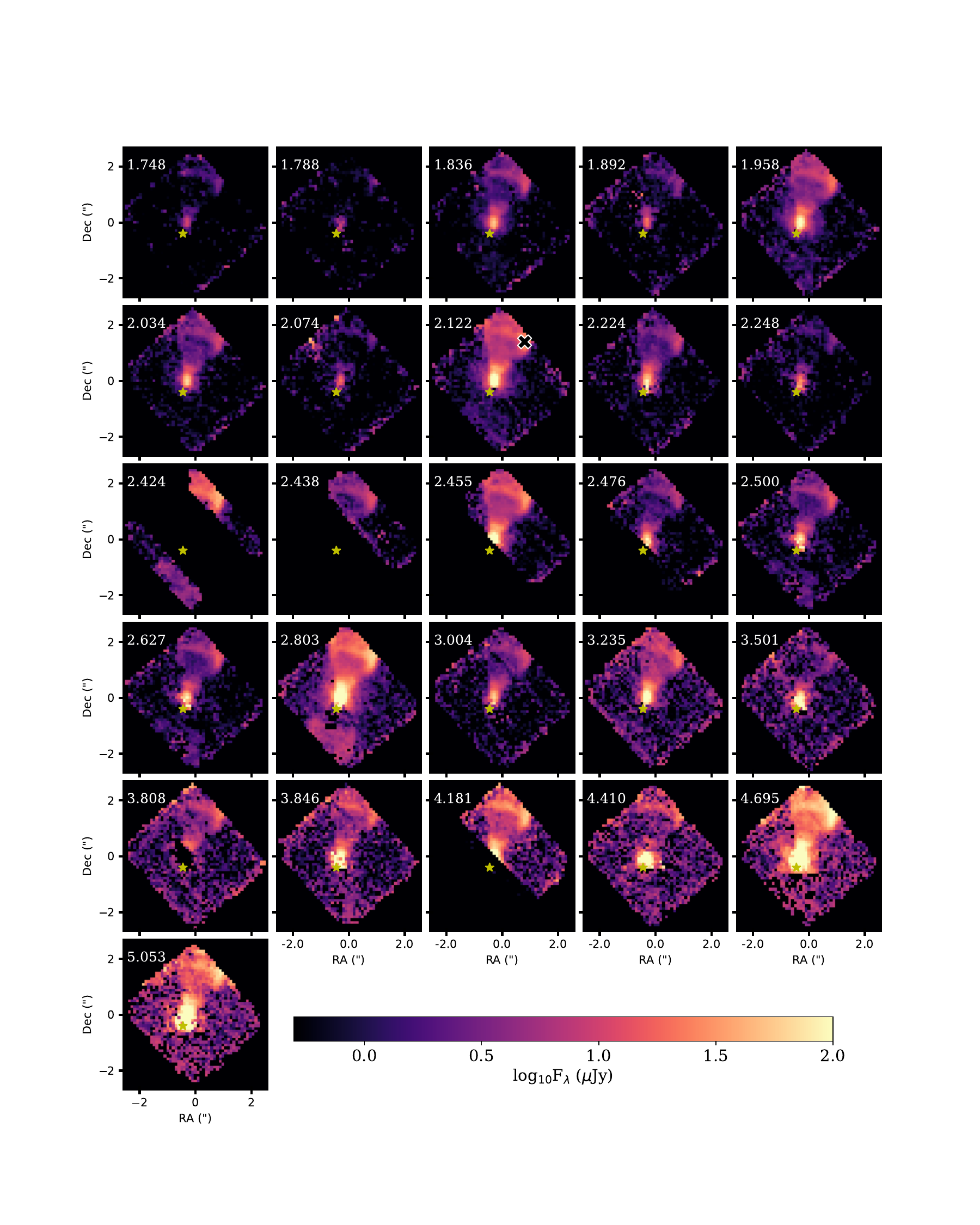}
\caption{Peak flux density maps of the detected \ce{H2} lines similar to Fig.~\ref{fig:full_iron2lines_moment0}. The black "x" with white borders in the 2.122 $\mu$m panel corresponds to the spaxel where the spectra in Fig.\ \ref{fig:full_H2lines_spectra} are taken.  } 
\label{fig:full_H2lines_moment0}
\end{figure*}


\subsection{[\ion{Fe}{II}] and \ce{H2} Detections}
\label{sec:iron_h2_detections}
We detect a total of 30 forbidden [\ion{Fe II}] emission lines to the north of the protostar (the blue-shifted region) between 1--5.3 $\mu m$ (see Sect.\ \ref{sec:atomicjet}). The spectra and peak line flux density maps (in units of $\mu$Jy) for all detected transitions are presented in Figures \ref{fig:full_iron2lines_spectra} and \ref{fig:full_iron2lines_moment0}, respectively. The spectra in Fig.\ \ref{fig:full_iron2lines_spectra} are extracted from a single spaxel (see black "x" in the 1.644 $\mu$m panel of Fig.\ \ref{fig:full_iron2lines_moment0}) where the continuum and scattered light contamination is at a minimum while still recovering the fainter lines. The peak flux density maps are shown in Fig.\ \ref{fig:full_iron2lines_moment0}, where the velocity of each map matches with the velocity of the peak flux density for each line shown in Fig.\ \ref{fig:full_iron2lines_spectra}. We find that this approach allows for the clearest distinction of the extended structure of the blue-shifted jet. We note that the 4.889 $\mu m$  [\ion{Fe}{II}] line is contaminated by strong CO fundamental (v$=$1--0) ro-vibrational emission closer to the protostar. As a workaround, the spectrum of this particular line is extracted from a spaxel at a larger offset (cf.\ the location of the "x" in the 4.889 $\mu m$ panels in Fig.\ \ref{fig:full_iron2lines_moment0}). The [\ion{Fe}{II}] detections include lines originating from the a2G, a4D, a4P, a2P, and a2H levels, with upper energies ranging from 2,430-26,055 K. The details of each line and the observed line ratio with respect to the 1.644 $\mu$m line are reported in Appendix \ref{App:detected lines}. 

In Figures \ref{fig:full_H2lines_spectra} and \ref{fig:full_H2lines_moment0}, we present the 26 \ce{H2} emission lines that we detect. Figure \ref{fig:full_H2lines_spectra} shows the 1D spectra for a bright spaxel in the north-east section of the \ce{H2} outflow (see the black "x" in the 2.122 $\mu$m map in Fig.\ \ref{fig:full_H2lines_moment0}).  The chosen spaxel is taken from one of the brightest regions of the \ce{H2} outflow. This location also benefits from being far from the protostar to minimize contamination from the bright scattered dust continuum. The continuum is stronger at wavelengths $>$ 2 $\mu$m compared to the 1–2 $\mu$m range, where most of the [\ion{Fe}{II}] lines are detected. As was done for [\ion{Fe}{II}], the flux density maps in Fig.\ \ref{fig:full_H2lines_moment0} are shown at the velocities given by the peak line intensities in Fig.\ \ref{fig:full_H2lines_spectra}. 
The \ce{H2} lines detected include the S, Q, and O rotational transition branches, with upper energies spanning from 6,000 K to 17,500 K. The details of each line and the observed line ratio with respect to the 2.122 $\mu$m line are reported in the Appendix \ref{App:detected lines}. 

\subsection{The Bipolar Atomic Jet}\label{sec:atomicjet}
\begin{figure*}
\centering
\includegraphics[width=\textwidth,trim=0cm 4cm 0cm 3cm, clip]{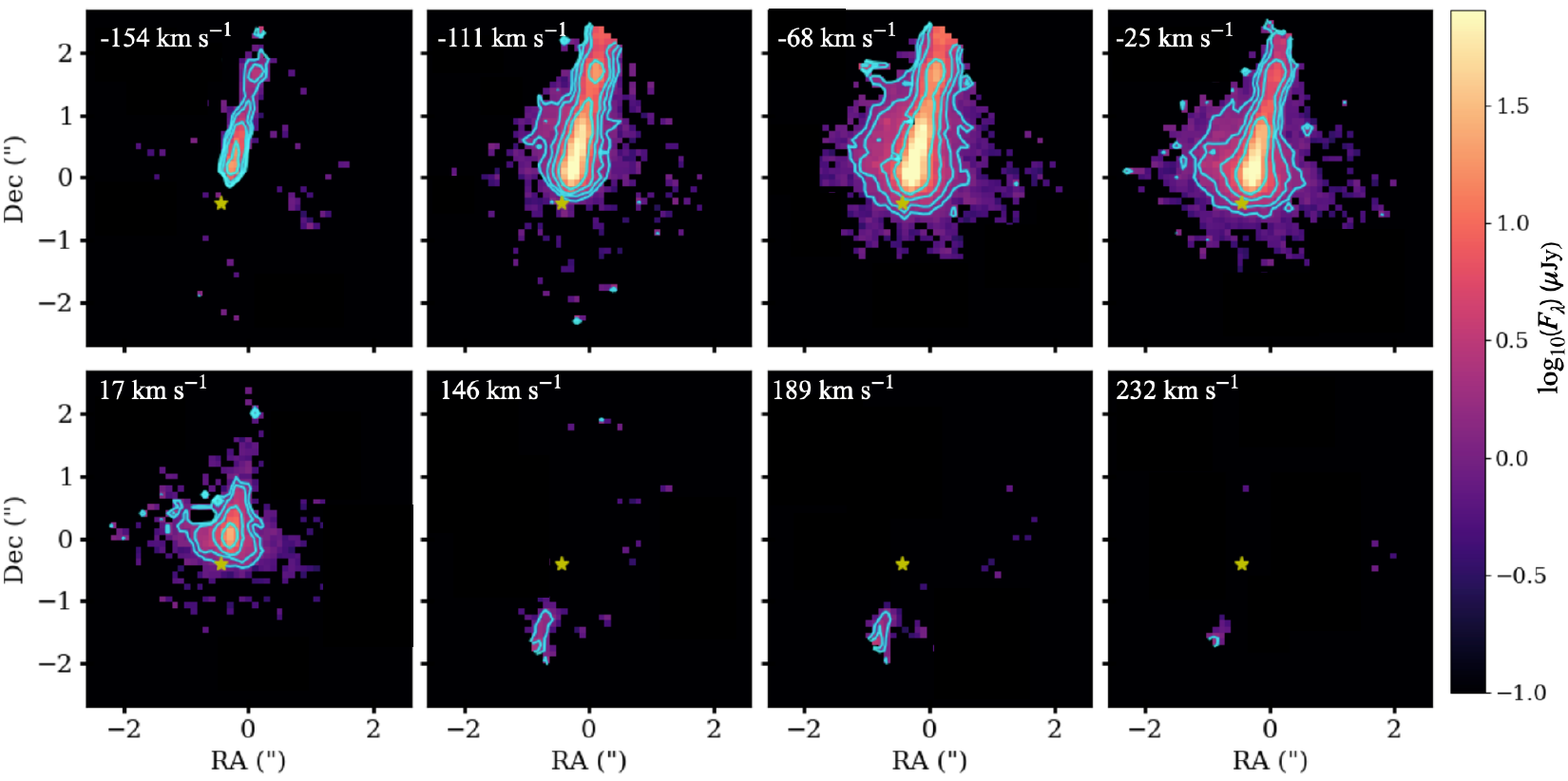}
\caption{Channel maps of the 1.644 $\mu$m [\ion{Fe}{II}] line. Contours are plotted at 1.5, 2.5, 5, and 15 $\mu$Jy.}
\label{fig:figfe2_at_diff_vels}
\end{figure*}
\begin{figure*}
\centering
\includegraphics[width=0.81\textwidth, trim=0cm 0cm 0cm 0cm,clip]{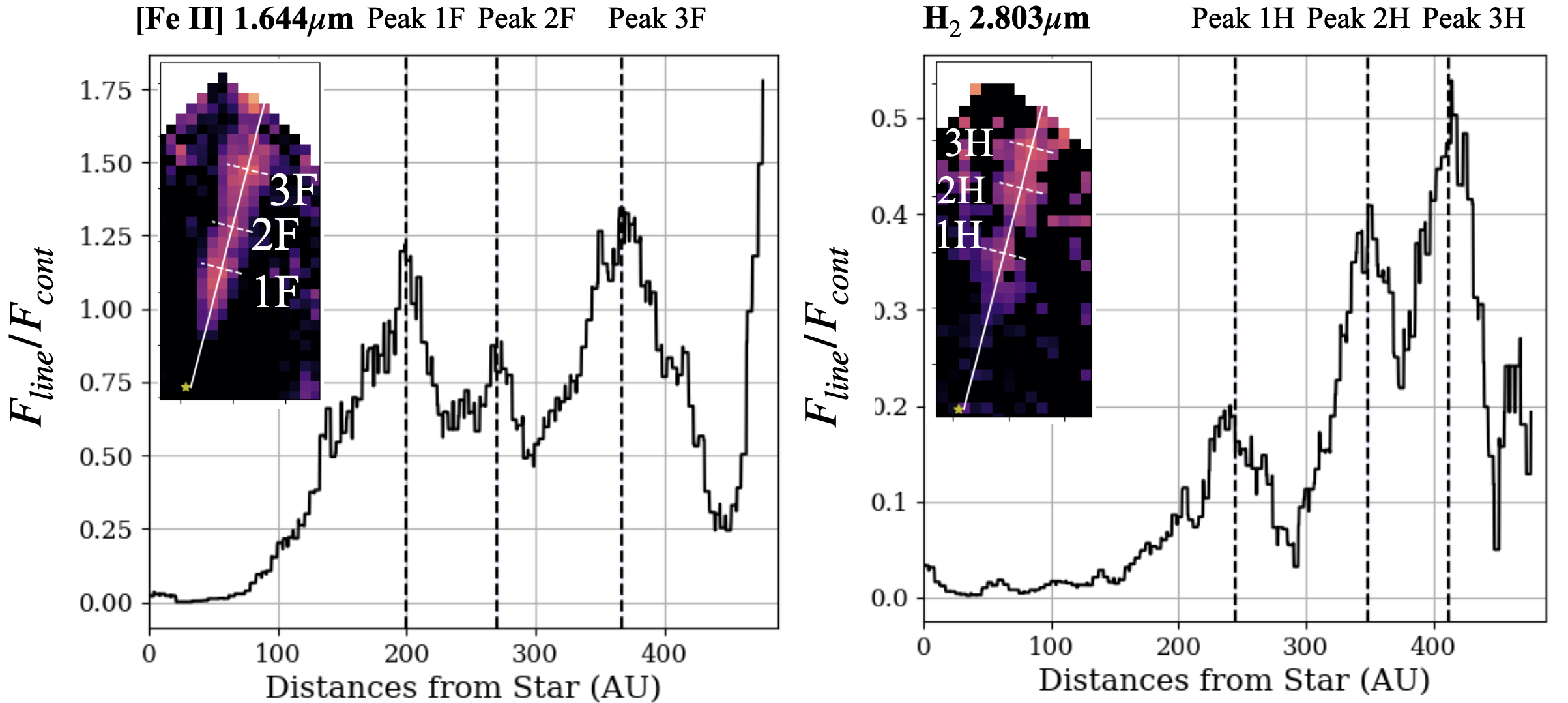}
\caption{Peaks in the [\ion{Fe}{II}] (Left) and \ce{H2} (Right) outflows.  The flux density profile of the highest velocity emission as a function of the deprojected distance along the jet's axis (-150 km s$^{-1}$ for [\ion{Fe}{II}], -110 km s$^{-1}$ for \ce{H2}) is shown normalized with respect to the dust continuum. The inset in each panel shows the flux density maps with color maps logarithmically spaced from 10$^{-2}$ to 10 $\mu$Jy.  The dashed vertical lines correspond to the peak (central) position of each peak determined by fitting Gaussian profile to the flux density profile (F$_{line}$/F$_{cont}$ vs distance). The same locations are indicated by the white dashed lines in the flux density map in the inset.}
\label{fig:figknots}
\end{figure*}

\begin{figure*}
\centering
\includegraphics[width=\textwidth,trim=0cm 3.5cm 0cm 3cm, clip]{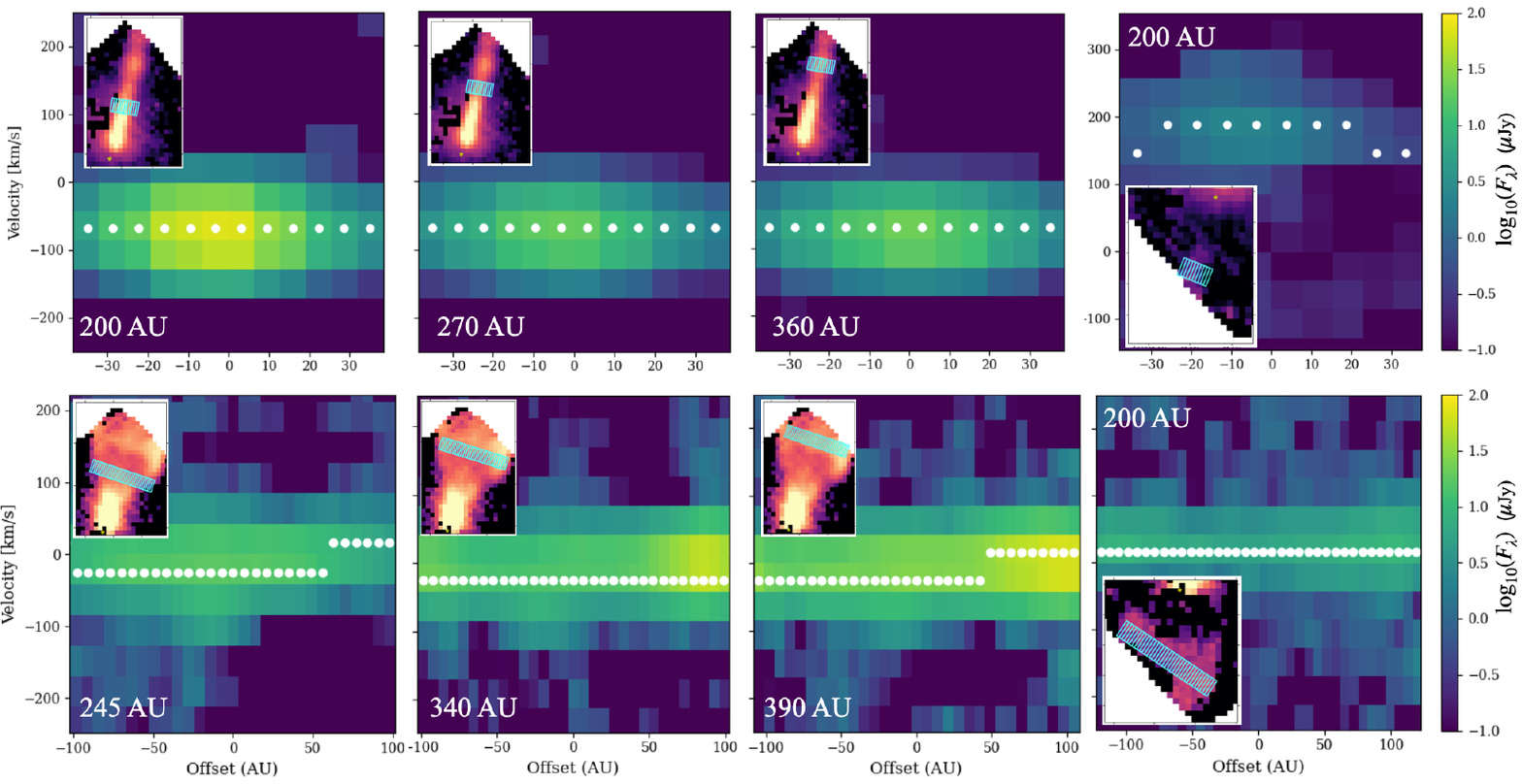}
\caption{PV diagrams of the 1.644 $\mu$m [\ion{Fe}{II}] (Top Row) and 2.803 $\mu$m \ce{H2} (Bottom row) lines at different locations in the bipolar jet. The distance offset from the jet axis is calculated by adopting a distance of 140 pc without de-projection since the slices are taken perpendicular to the jet.  The de-projected offsets are shown in the bottom left of each panel. In each panel,  the integrated flux density or moment-0 maps are shown in the inset along with the region that the PV diagram was taken which has a width of 4-spaxels.
} 
\label{fig:figPVdiagrams}
\end{figure*}

We investigate the spatial and velocity structure of the bright 1.644 $\mu m$ [\ion{Fe}{II}] line to characterize the atomic jet emanating from TMC1A. \cite{harsono2023_tmc1a} revealed a bright blue-shifted atomic jet to the northern side of the protostar in TMC1A. In addition to this blue-shifted component, we detect a dimmer red-shifted component of the jet on the southern side of the protostar in the 1.257, 1.644 \& 1.803 $\mu$m lines, the first indication that the atomic jet in TMC1A is bipolar.

 In Figure \ref{fig:figfe2_at_diff_vels}, we show the flux density maps of the 1.644 $\mu$m [\ion{Fe}{II}] line across several velocity bins, overlaid with contours at several flux levels with signal-to-noise greater than 3. At 1.644 $\mu$m, the width of each velocity bin is approximately 40 km s$^{-1}$. The flux density of the blue-shifted component peaks at around -70 km s$^{-1}$, with emission detected at velocities upwards of approximately -150 km s$^{-1}$. The red-shifted side, although it is 10-30$\times$ dimmer than the blue-shifted side, has its peak flux density at a velocity of $\sim$190 km s$^{-1}$ with tentative evidence of high-velocity components close to $\sim$230 km s$^{-1}$. However, the highest velocity components of the red-shifted outflow are only found close to the edge of the IFU FOV, and need confirmation with additional observations.

Due to the protostar being offset from the center of the FOV of our observations, more of the blue-shifted jet is within the FOV, and because the blue-shifted jet is substantially brighter, we detect spatially distinct features in the flux density maps. The structure of the atomic jet is characterized by a distinctive peak in the flux density profile coinciding with the apparent jet axis, flanked on either side by a sharp decrease. To highlight the location of the peaks, we take 1D slices of the flux density maps along the axis of the [\ion{Fe}{II}] jet and normalize the flux density at the highest velocity component of the jet ($F_{line}$) to the continuum flux density at that velocity ($F_{cont}$). This shows the flux density of the fastest component of the jet for the 1.644 $\mu$m line in terms of its strength in comparison to the continuum and makes it easier to determine the location of the peaks.

Within the FOV of these observations, we identify three distinct intensity peaks visible in the brightest [\ion{Fe}{II}] lines (e.g., 1.257, 1.644, 1.810 $\mu m$) on the northern side of the atomic jet, as shown in Figure \ref{fig:figknots}. For brevity, we only show the 1.644 $\mu$m line and note that the location of the peaks does not change significantly in the other lines. We determine the location of these peaks by fitting Gaussian profiles to the normalized 1D flux density profiles. The distances of each peak are reported in AU from the central star and de-projecting each point along the jet adopting an inclination with respect to the plane of the sky of $\sim 35^{\circ}$ for the jet, based on the disk inclination of $\sim 55^{\circ}$ (where $90^{\circ}$ would be edge-on) \citep{harsono2021resolved}.
 
The first peak (1F) appears as as an extended structure with a maximum in its flux density profile at a de-projected distance of $\sim 200$ AU (see Figure \ref{fig:figknots}).
The second emission feature (2F) is only visible in the brighter [\ion{Fe}{II}] lines and is considerably smaller than the other two, located at a distance of approximately 270 AU from the protostar. The third emission feature (3F) is observed in several [\ion{Fe}{II}] lines and is located close to the edge of the FOV at $\sim$370 au from the protostar. 
Adopting a de-projected line-of-sight tangential velocity $v_{\rm gas}=v_{\rm obs}/\sin (35^{\circ}$) = -150 km s$^{-1}$/sin(35$^{\circ}$)$\approx$ -260 km s$^{-1}$, the location of these peaks correspond to dynamical timescales ($t_{dyn} = \frac{d}{v_{gas}}$) of 4, 5, and 7 years, respectively, implying a $\sim$1-2 year dynamical time difference between adjacent peaks.

We employ position-velocity (PV) diagrams to examine velocity profiles perpendicular the jet axis for each distinct emission feature in the bright 1.644 $\mu m$ line (Figure \ref{fig:figPVdiagrams}, top row). Utilizing the \href{https://pvextractor.readthedocs.io/en/latest/}{Astropy PV Extractor}, we construct PV diagrams by defining a line perpendicular to the jet axis with a width of 3 pixels, and the velocity space is computed using the rest wavelength of the line. Solid white dots overlaid on the diagrams represent velocities at the peak line flux density. At this spectral resolution, no clear indication of a velocity gradient can be discerned. Theoretically, protostellar outflows are expected to rotate, and indeed there is evidence to support it \citep[e.g.][]{franketal14_ppvi}. We would expect the rotation to show up as an increasing or decreasing velocity gradient perpendicular to the jet axis, as seen in some sources \citep[e.g., DG Tau,][]{zapata2015kinematics, de2020alma}. In these observations, the velocity at peak flux density remains constant across the jet axis within the velocity resolution and does not change significantly from one peak to the next.

\subsection{The Molecular Jet and Wider Angle Outflow}
\label{sec:molecular_jet_and_outflows}
\begin{figure*}
\centering
\includegraphics[width=\textwidth,trim=0cm 4cm 0cm 4cm, clip]{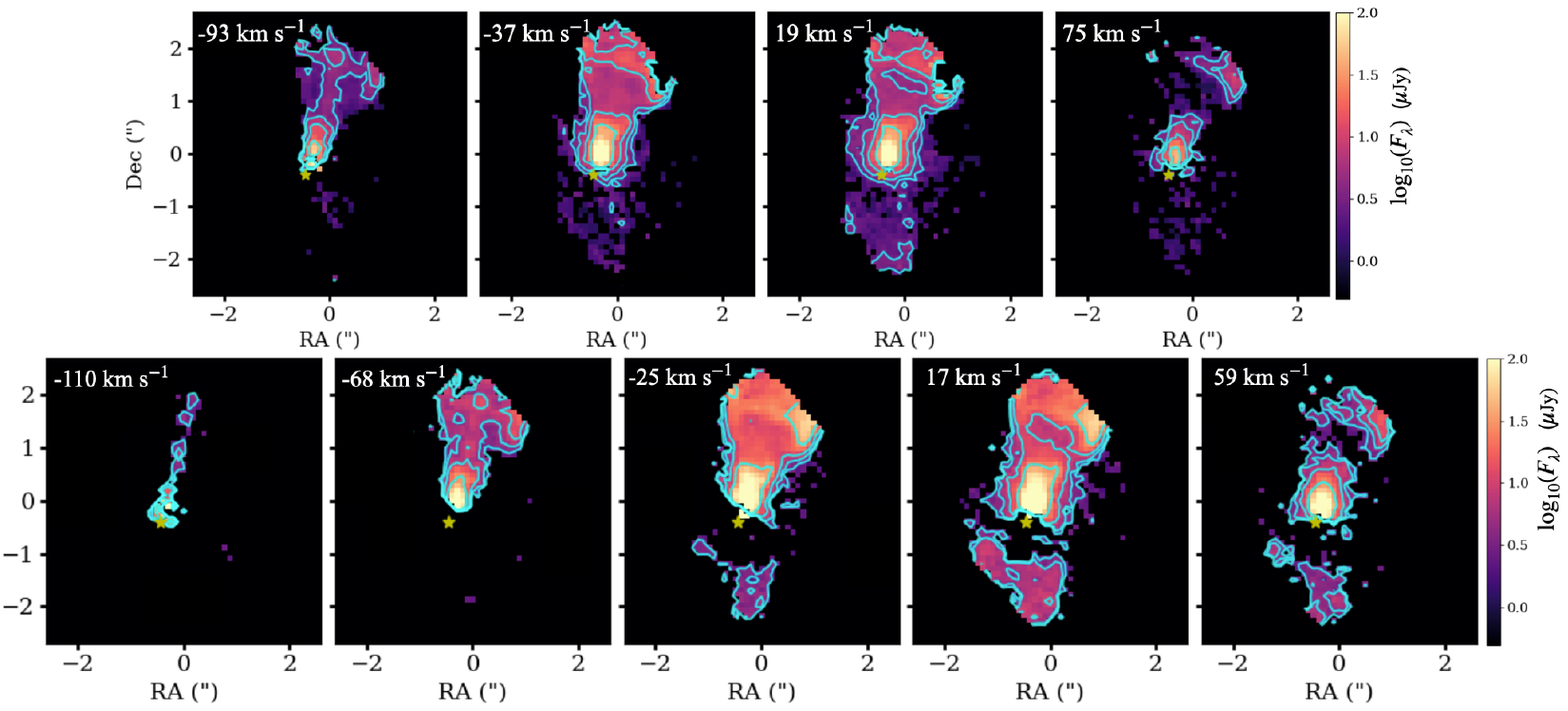}
\caption{Channel maps of the 2.122 (top panels) \& 2.803 $\mu$m (bottom panels) \ce{H2} lines. Contours are plotted at flux levels of 3 ,5,10, and 30 $\mu$Jy shown by the cyan lines. 
} 
\label{fig:h2_velocity_bins}
\end{figure*}


\begin{figure*}
\centering
\includegraphics[width=0.82\textwidth]{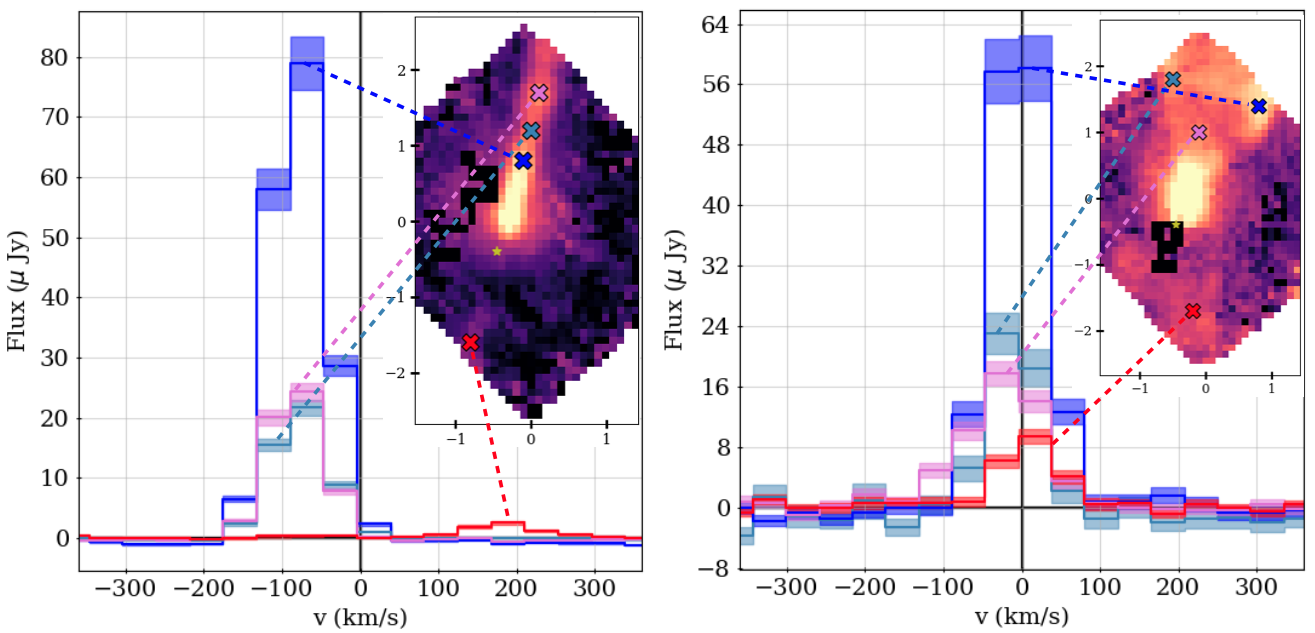}
\caption{Line profiles of the 1.644 $\mu$m [\ion{Fe}{II}] (Left) and 2.803 $\mu$m \ce{H2} (Right) lines at different locations.  The spectra are color-coded and correspond to the locations marked by "x" in the moment-0 maps shown in the insets. The spectra are extracted in different locations for the [\ion{Fe}{II}] and \ce{H2} lines to highlight particular characteristics. In the left panel, these correspond to the 3 brightness peaks along the blue-shifted jet axis (blue, steel blue, purple) and red-shifted jet (red). In the right panel, the spectra correspond to the central spine with higher velocity emission detected (purple), the left (steel blue) and right (blue) wings of the ring-like structure and the red-shifted outflow (red).} 
\label{fig:figspectral_diffs}
\end{figure*}

As discussed in \cite{harsono2023_tmc1a}, extended molecular hydrogen (\ce{H2}) emission is detected towards TMC1A with a nearly conical shape that appears to enclose the blue-shifted [\ion{Fe}{II}] jet. Similar to our analysis of the [\ion{Fe}{II}] lines, we investigate the morphology of two of the bright \ce{H2} lines (2.122 \& 2.803 $\mu$m) in detail. For both lines, the structure of the outflow depends on velocity. To examine the kinematic structure of the \ce{H2} emission, we plot flux density maps for each velocity channel with significant \ce{H2} emission in Figure \ref{fig:h2_velocity_bins}. In the highest velocity bins of both 2.122 \& 2.803 $\mu$m  lines, there is evidence for a collimated structure coincident with the axis of the [\ion{Fe}{II}] jet. Beyond the jet axis, there is little to no emission at higher velocities (leftmost panel in the top and bottom row of Fig.\ \ref{fig:h2_velocity_bins}). The line-of-sight velocity for the 2.122 $\mu$m \ce{H2} line reaches -90 km s$^{-1}$, and -110 km s$^{-1}$ for the 2.803 $\mu$m, corresponding to -160 km/s and -190 km/s respectively after de-projection. This lends evidence for a molecular jet component in addition to the atomic jet in TMC1A. 

The slower velocity components ($\sim$ -30 - +20 km/s) of the \ce{H2} lines make up the wider angle ("shell") outflow. In comparison to the atomic jet, the width of the \ce{H2} outflow is approximately 240 AU at a deprojected distance of 350 AU (2") from the star while the [\ion{Fe}{II}] has a width of $\sim$70 AU at this distance. There is also low velocity \ce{H2} emission present on the red-shifted side of the outflow, demonstrating that the \ce{H2} outflow is also bipolar, similar to the atomic jet. The southern part of the \ce{H2} outflow is also dimmer than the northern part by more than a factor of 10. However, unlike the narrow atomic jet, the red-shifted component of the \ce{H2} outflow is wider (a width of $\sim$ 250 AU at 150 AU from star compared to $\sim$50 AU for the redshifted [\ion{Fe}{II}] outflow at the same distance) and only apparent at lower velocities. The redshifted \ce{H2} outflow is also wider than its northern counterpart at the same distance ($\sim$250 AU compared to $\sim$ 110 AU at 150 AU from the star). 

In the northern \ce{H2} outflow, a ring-like structure is observed that exhibits a brightness asymmetry where the right side is 3-4$\times$ brighter than the left side (see Figure \ref{fig:figspectral_diffs}). The right side includes more red-shifted flux compared to the left, which shows little emission at $\sim$ 60 km s$^{-1}$ (see Figure \ref{fig:h2_velocity_bins},\ref{fig:figspectral_diffs}). This may indicate an unresolved velocity gradient associated with the asymmetry.

We further examine the velocity structure using PV diagrams of the 2.803 $\mu$m line taken perpendicular to the jet/outflow axis. As with [\ion{Fe}{II}], in Fig.\ \ref{fig:figPVdiagrams} (bottom row), solid white dots mark the velocity at peak line emission, which are observed at approximately -30 km s$^{-1}$ and 20 km s$^{-1}$ for the blue-shifted and red-shifted components, respectively. In the first and third columns of the PV diagrams for \ce{H2}, a transition in peak velocity from -30 km s$^{-1}$ to 20 km s$^{-1}$ is observed at the location of the brightness asymmetry in the ring of the \ce{H2} shell. However, this shift is not discernible at an intermediate location (the second column) due to marginal differences in the peak flux density between the blue-shifted and red-shifted velocity bins. Nonetheless, the intensity of the PV diagram still shows an increasing flux density towards the first red-shifted velocity bin at this location.

To provide more detailed insights, we plot the 1D spectra at individual spaxels for the 2.803 $\mu$m \ce{H2} line in the right panel of Figure \ref{fig:figspectral_diffs}. The left-side (cyan) of the outflow covers $\sim$3 spectral channels (-70 - +20 km s$^{-1}$) with the peak emission observed at $\sim$ -30 km s$^{-1}$. This is in contrast to the right side of the outflow which covers $\sim$4 channels (-70 - + 60 km s$^{-1}$) and peaks at $\sim$20 km s$^{-1}$ (light red). However, in the southern outflow, the emission includes $\sim$3 spectral channels which lean towards more red-shifted velocities ($\sim$-30 - +60 km s$^{-1}$) (dark red), presumably because the spectra at this location traces the red-shifted outflow. Toward the central axis of the northern outflow (dark blue), the line appears broader ($\sim$5 channels wide) and shows emission at a notably higher blue-shifted velocity, approximately -110 km s$^{-1}$, corresponding to the jet.

Using the same method as for the [\ion{Fe}{II}] 1.644 $\mu$m line,  peaks in the flux density of  the \ce{H2} jet are determined by constructing a normalized flux density profile along the jet's axis at the highest velocity emission (-110 km s$^{-1}$) of the 2.803 $\mu$m line and normalize by the continuum (right panel, Figure \ref{fig:figknots}). We find that there are similarly three peaks (1H, 2H, 3H) in the blue-shifted \ce{H2} jet. However, each peak is located at a farther distance from the protostar with respect to the observed [\ion{Fe}{II}] peaks (see Sect.\ \ref{sec:atomicjet}). With a deprojected velocity of $\sim$-190 km s$^{-1}$, the three identified peaks in the molecular jet are located at de-projected distances of 245, 350, and 410 AU,  corresponding to dynamical timescales of 6, 9, and 10 years. Notably, the locations of the \ce{H2} peaks are found in the brightness dips of the atomic jet (left panel, Figure \ref{fig:figknots}) suggesting they are situated in between the [\ion{Fe}{II}] peaks, although higher spatial resolution observations are needed for confirmation.

\subsection{Extinction of the Jet} 
\label{sec:extinction}

\begin{figure}
\centering
\includegraphics[width=0.5\textwidth, trim=1cm 2cm 1cm 2cm]{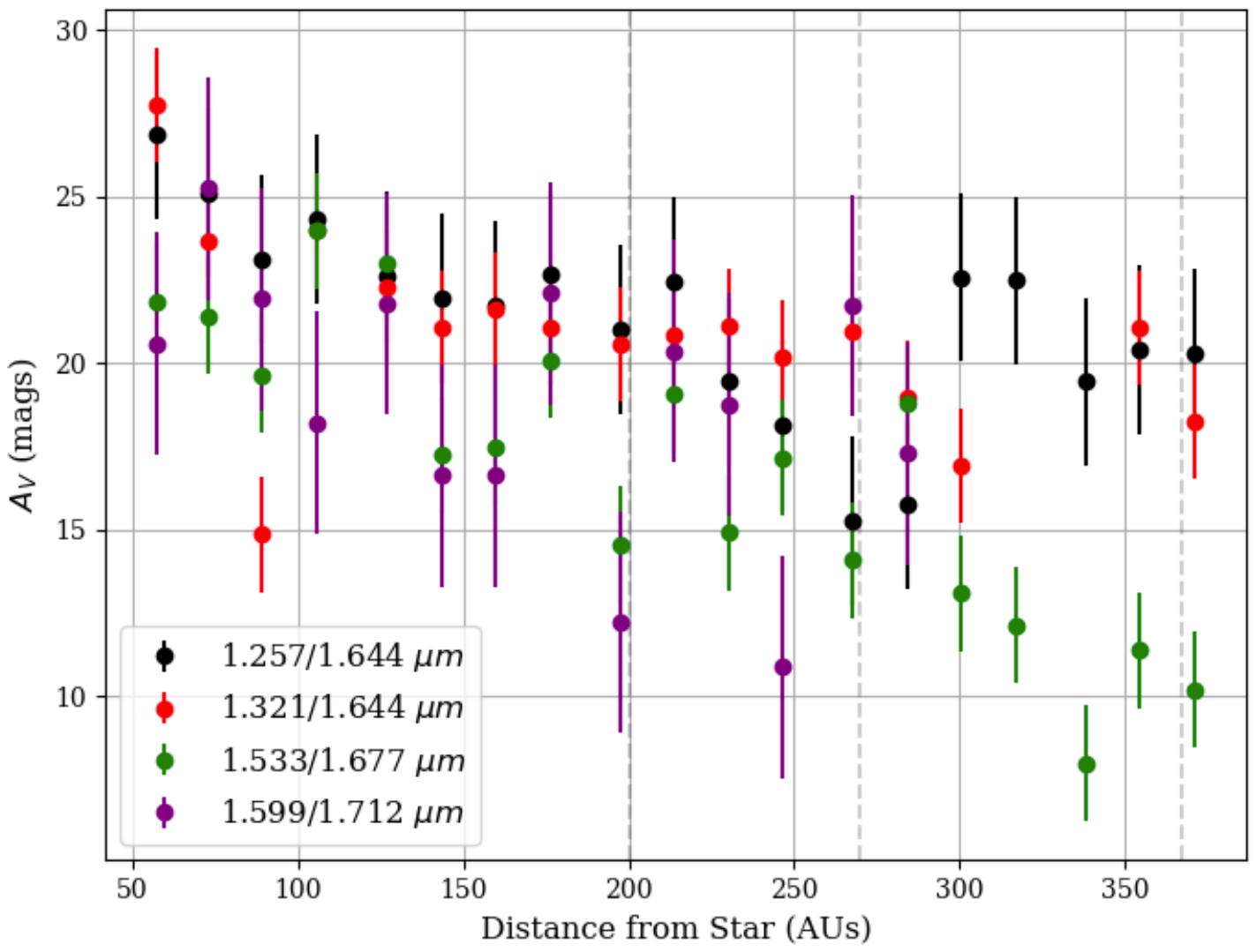}
\caption{Visual extinction ($A_V$) towards the jet determined from the observed [\ion{Fe}{II}] line ratios with respect to their intrinsic ratios. We calculate the line ratios at peak line flux for spaxels along the northern (blue-shifted) jet axis. We use the ratio at peak line flux since the lines are only marginally resolved in wavelength (3--4 channels). The observed trends are consistent with the values derived from taking line ratios with zeroth-moment maps which have higher uncertainty due to spaxels that contain faint emission.}
\label{fig:figextinction}
\end{figure}

The extinction of radiation is a fundamental measure representing the absorption and scattering of light by foreground dust, causing a reddening effect. Accurate determination of the extinction towards the jet is crucial to knowing the actual brightness of several lines used for constraining physical conditions, such as excitation temperatures and densities. We estimate the extinction towards the jet using the observed [\ion{Fe}{II}] ratios. The intensity ratio of optically thin lines of the same atom that originate from the same upper level does not depend on the temperature or density of the emitting gas, but rather the ratio of their Einstein A-coefficients via $I_1/I_2 = A_1 \nu_1/(A_2 \nu_2$) \citep{gredel1994near,giannini2015empirical,Erkaletal2021}, where $\nu_1$ and $\nu_2$ are the frequencies of the two lines. A difference between an observed line ratio value and the intrinsic ratio can therefore be attributed to line-of-sight extinction. Since we detect several transitions of [\ion{Fe}{II}] in the jet of TMC1A, we use them to probe extinction. The observations of TMC1A are sufficiently good that we can estimate the extinction of the jet using four separate line ratio pairs.
We adopt a similar methodology to \citet{Erkaletal2021} for determining the extinction of the jet by comparing observed line ratios to the intrinsic ratios tabulated in  \citet{giannini2015empirical}, and then using the parameterized extinction curve from \cite{cardelli1989relationship} to determine the visual extinction, $A_{\rm V}$, in magnitudes.

We use Equation 21.1 of \cite{draine2010physics}, which describes how the extinction at a given wavelength, $A_{\lambda}$, reduces the observed flux $F_{obs}(\lambda) = F_{int}(\lambda) \times 10^{-0.4 A_{\lambda}} = F_{int}(\lambda) \times 10^{-0.4 \beta_{\lambda}A_V}$,
where $F_{int}$ is the intrinsic flux and $\beta_{\lambda}=A_{\lambda}/A_{\rm V}$. The $\beta$ terms are determined using the extinction curve from \cite{cardelli1989relationship}, which demonstrated that extinction can be determined solely based on the parameter R$_V$ that describes the slope of the extinction at visible wavelengths. We adopt R$_V$=5.5 as in \cite{mcclure2019carbon}. Given the well-established intrinsic flux ratios for certain [\ion{Fe}{II}] lines \citep{giannini2015empirical}, it is useful to take the ratio of the previous equation at two of these specific wavelengths to then solve for $A_{\rm V}$ \citep{giannini2015empirical}:
\begin{equation}
    A_V=-2.5\log_{10}\left(\frac{F_{obs, \lambda_1}/F_{obs, \lambda_2}}{F_{\text{int}, \lambda_1}/F_{\text{int}, \lambda_1}}\right)(\beta_{\lambda_1} - \beta_{\lambda_2})^{-1},
\end{equation} 
where $\beta_{\lambda_1}$ and $\beta_{\lambda_2}$ represent the ratio of $A(\lambda)/A(V)$ calculated from the extinction curve of \cite{cardelli1989relationship} (their eq.\ 1). To estimate the uncertainty in the extintion, we propagate uncertainties in the standard way for the intrinsic ratio ($F_{\text{int}, \lambda_1}/F_{\text{int}, \lambda_2}$) using the standard deviations reported in \cite{giannini2015empirical}, and the observed peak flux ($F_{obs, \lambda_1}$, $F_{obs, \lambda_2}$). 

Due to the lines spanning only 3-5 velocity bins, we determine the ratio of observed intensities by taking the ratio of the maximum flux densities of each line along the jet's axis at several different locations/spaxels. The intrinsic ratios of the 1.257/1.644, 1.321/1.644, 1.533/1.677, and 1.599/1.712 $\mu m$ pairs used in this study are taken from \cite{giannini2015empirical}, and are taken to be 1.1, 0.32, 1.25, and 3.6, respectively. However, the intrinsic line ratio can vary as a result of how the transition probabilities are calculated. For example, the 1.257/1.644 ratio can vary between 0.98-1.49 \citep{rodriguez_2004_h2_fe_AGN, smith_fe_pcygni_2006, giannini2015empirical} and more recently an even higher value of 2.6 has been proposed \citep{rubinstein_extinction_2021}. For the same observed flux ratio, a difference in the intrinsic ratio of 0.98 to 1.49 corresponds to a difference in $A_{\rm V}$ of $\sim$4 magnitudes while 0.98 to 2.6 increases the difference to $\sim$9. In TMC1A, the observed line ratios for the 1.257/1.644 $\mu m$ pair range from 0.06-0.2 (values are lower near the protostar). This gives $A_{\rm V}$ of 14-26 (higher $A_{\rm V}$ near the protostar) for an intrinsic ratio of 0.98 and 23-35 for an intrinsic ratio of 2.6. This uncertainty in the intrinsic line ratio results in a substantial uncertainty in the inferred extinction values.

In Figure \ref{fig:figextinction}, we plot the derived extinction as a function of the de-projected distance along the jet's axis. The range of $A_{\rm V}$ values span from 10-30 magnitudes with uncertainties of approximately $\pm$ 2-3. Only spaxels with S/N $>3$ are shown. We find significant discrepancies in our derived values of extinction when using the less bright [\ion{Fe}{II}] line ratios (i.e.\ 1.533/1.677 and 1.599/1.712). The reason for this discrepancy remains unclear. Nevertheless, across all four line ratio pairs, larger extinction measurements are consistently observed closer to the protostar, gradually decreasing as we move along the jet axis. These findings align with those of \cite{connelley2010near} for TMC1A (IRAS 04365+2535), where a continuum extinction of $A_{\rm V}$=30$^{+9}_{-30}$ was reported.

We adopt the same analysis for the dimmer southern (red-shifted) jet. Despite that the needed [\ion{Fe}{II}] lines are only detected in a few spaxels in the 1.257 $\mu m$ line, we can estimate the extinction towards the red-shifted jet. For two of the brightest spaxels showing emission at 1.257 $\mu m$ and 1.644 $\mu m$, the extinction of the red-shifted jet is estimated to be $A_{\rm V} \sim 13.8\pm2.5$ through the 1.257/1.644 ratio (F$_{1.257}$/F$_{1.644}$=0.24) at the brightest spaxel of the red-shifted side and $A_{\rm V} \sim17.4\pm2.5$ for the next brightest spaxel (F$_{1.257}$/F$_{1.644}$=0.17). Both red-shifted spaxels are located at a de-projected distance of approximately 200 AU from the protostar. At the same distance in the blueshifted jet, the flux ratio is F$_{1.257}$/F$_{1.644}=0.09-0.12$ corresponding to $A_{\rm V}=21-23$  Therefore, the extinction we measure for the red-shifted side is comparable to, or even slightly less, than the blue-shifted jet, despite having a much lower flux density. 

\section{Discussion}
\label{sec:discussion}
\subsection{Asymmetries in the Bipolar Outflows}

For the first time, we find clear evidence of a red-shifted atomic [\ion{Fe}{II}] jet in TMC1A moving at a deprojected velocity of $\sim 330$ km/s to the south of the protostar. In contrast, most of the emission of the blue-shifted atomic jet peaks at a deprojected velocity of $\sim -120$ km/s, but has emission detected at $\sim$ -260 km s$^{-1}$. The detection of both blue- and red-shifted jets indicates the presence of an embedded bipolar jet. In addition, the southern, red-shifted, component of the \ce{H2} outflow is also revealed. 
The southern outflow in both the [Fe II] jet and the \ce{H2} outflow is 10-30$\times$ fainter than the northern counterpart (see Figure \ref{fig:figspectral_diffs}). Additionally, the \ce{H2} outflow exhibits a brightness asymmetry about the axis of the jet in a "ring"-like emission feature, where one side of the ``ring'' is 3-4$\times$ brighter than the other. Below, we address the observed asymmetries and their implications.

\subsubsection{Northern and Southern Outflows}\label{sec:disc_bipolar_asym}

Given the observed intensity and velocity differences in the bipolar outflow components, it is important to determine whether these differences are intrinsic, e.g., in the strength or power of the outflows (which we relate to the brightness) or due to external factors such as a non-uniform ambient medium. In the case of TMC1A, our results suggest that differential extinction (from one side of the bipolar outflow to the other) is not the cause of the observed intensity differences: the extinction found in the blue-shifted component of the bipolar jet is comparable to the red-shifted component (see Sect.\ \ref{sec:extinction}). If the asymmetry was caused by differential extinction, a brightness difference of 10-30$\times$ would mean the southern outflow would have a higher extinction on the order of $\Delta A_{\lambda}$ = 2-4 magnitudes. Assuming this \textit{difference} in extinction is the same for \ce{H2}, then the same argument would apply since it is also 10-30$\times$ fainter in the south. However, the extinction measured in the southern outflow using the [Fe II] lines is comparable, if not slightly less than the northern outflow. This conclusion is independent of the uncertainties in the intrinsic line ratios, because the observed ratio of the 1.257 $\mu$m and 1.644 $\mu$m lines in the southern (red-shifted) jet is comparable to, or somewhat higher than, the corresponding location in the northern (blue-shifted) jet.

A more extincted northern outflow could be caused by a non-uniform line-of-sight gas and dust distribution. Further evidence of the bipolar asymmetry can be seen in the sub-millimeter CO $J=$2-1 (230.5 GHz) emission, which shows the pronounced blue-red side asymmetry, especially in the highest velocity bin (beyond 9 km/s from the line center, see Fig.\ 13 of \cite{aso2021}). The highly collimated CO outflow is essentially one-sided with only the blue-shifted component, which is less affected by dust extinction in the sub-millimeter compared to the infrared. Additional evidence of the asymmetry can be found in other shock tracers such [O I] at 63.2 $\mu$m, and OH at 84.6 $\mu$m, where detected emission was only found towards the north of the protostar \citep{wish_survey_karska_2013}. The similar asymmetric trend is also seen in TMC1A in \ce{^{12}CO} $J=$14-13 \citep{wish_survey_karska_2013}, \ce{^{13}CO} $J=$6-5, $J=$3-2 emission \citep{Yildeiz_APEX_2015} and HCO$^+$ 4-3 emission in the Leiden Observatory Single-dish Sub-mm Spectral Database of Low-mass YSOs (LOMASS) \citep{LOMASS_carney_2016}.

At lower velocities ($\lesssim$9 km s$^{-1}$), the northern and southern molecular outflow appears nearly symmetric \citep{aso15, aso2021}. Furthermore, the large-scale protostellar envelope dust emission around TMC1A, as observed at 450 and 850 $\mu m$ with the James Clerk Maxwell Telescope's (JCMT) Submillimetre Common-User Bolometer Array camera (SCUBA) \citep{SCUBA_DI_francesco_2008}, shows no evidence of such asymmetry, other than diffuse emission to the north of TMC1A that may be associated to the surrounding molecular cloud. The lack of asymmetries at slower velocities (i.e.\ closer to systemic), and a largely symmetric dusty envelope further supports that the asymmetry observed in the [\ion{Fe}{II}] jet and \ce{H2} emission between the northern and southern sides is the result of an intrinsic difference, rather than, e.g., differential extinction.

In TMC1A, the spatially resolved blue-red asymmetry in the otherwise heavily extincted [\ion{Fe}{II}] and \ce{H2} outflows can be resolved thanks to the unrivaled sensitivity of the instrument. Follow-up {\it JWST} observations that map both sides of the outflow at larger distances from the protostar, beyond the FOV of our observations, and where extinction should be less, would be invaluable in confirming these results. Furthermore, a larger {\it JWST} survey of outflows towards more embedded protostars (Class 0-I) in different environmental conditions (inclination, extinction, etc.) will shed light on how common bipolar asymmetries are in younger sources. Indeed, blue-red side asymmetries have been observed in a growing number of older Class II sources \citep[e.g.][]{wowitas2002, melnikov2023}. Our observations of TMC1A indicate that this asymmetry can already be present in the younger Class I phase.

\subsubsection{The Ring-Like Structure in the \ce{H2} Outflow}\label{sec:disc_ring}

In many of the \ce{H2} detections presented, the molecular outflow exhibits a ring-like structure on the northern side, displaying a clear brightness asymmetry about the jet axis. The right side of this ring is 3-4 times brighter than the left side (see Fig.\ \ref{fig:figspectral_diffs}). Moreover, this left-right asymmetry is evident in the velocity distribution (Fig.\ \ref{fig:h2_velocity_bins}), where the right side is more red-shifted than the left side, indicating a large-scale velocity gradient from one side of the ring to the other. This trend is further supported in the PV diagrams (see Figure \ref{fig:figPVdiagrams}) where the velocity at peak intensity is more red-shifted in the region corresponding to the right side of the ring-like structure (-25 km s$^{-1}$ on left to $\sim$ +20 km s$^{-1}$ on right). The observed structure resembles the partial ring-like structure observed close to the protoplanetary disk of TMC1A in SO $N_{\rm J}=5_6-4_5$ in \cite{harsono2021resolved}, with a stronger and more red-shifted right side compared to the left. Intriguingly, the right side being more red-shifted than the left has the same directional sense as the rotation of the protoplanetary disk and CO molecular outflow in TMC1A \citep{aso15,bjerkeli2016, harsono2021resolved, aso2021}. That said, the \ce{H2} is likely interacting with an asymmetric surrounding medium, which could produce the observed left-right asymmetry in both brightness and line-of-sight velocity \citep[e.g.][]{De_colle_velocity_asym_2016}.


A better understanding of the left-right asymmetry may come from a comparison of the observed \ce{H2} ring-like structure to other molecular lines. \cite{harsono2021resolved} suggested that the SO emission originates from the warm and dense inner envelope close to the outflow cavity wall. At scales of a few 10s of au, both sub-mm HCN $J=3-2$ and \ce{HCO^{+}} $J=3-2$ emission also exhibit similar left-right asymmetry as SO and \ce{H2}. On larger scales ($\pm \sim 5-10$\arcsec), the sub-millimeter C$^{18}$O $J=2-1$ emission traces the flattened envelope, and shows an elongated red-shifted component toward the northwest (see Fig.\~4 of \citealt{aso15}) which may be associated with the infalling envelope. As mentioned in Sect.\ \ref{sec:disc_bipolar_asym}, the large-scale dusty envelope also shows a diffuse sub-mm emission along the same direction as the \ce{C^{18}O} emission at scales greater than 10\arcsec ($\gtrsim$ 1,400 AU) from the protostar. This connection may potentially point to a "feeding arm" that is actively adding mass from the molecular cloud to the protostar and preferentially interacting with the northern outflow, but more so where the \ce{H2} is brighter. In this scenario, the interaction between the outflow and the non-uniform ambient medium gives rise to the left-right asymmetry.

Finally, it is worth pointing out that the \ion{H}{I} and \ion{He}{I} emission \citep[see][]{harsono2023_tmc1a} is brighter on the left side (opposite of the bright \ce{H2} emission on the right side). The emission in both of these atomic lines is dominated by the scattered emission from the unresolved central protostar. The left-right brightness asymmetry could therefore simply arise from radiative transfer effects where scattering dominates the left side of the observed structure where the line-of-sight could go through the outflow cavity while the right side is the line of sight through the edge of the outflow cone providing higher column of gas and dust.

\subsection{Atomic and Molecular Jets in the Class I Stage}
TMC1A is classified as a Class I protostellar source. It is in an intermediate stage between Class 0 and II sources, possessing both a protostar and disk system while still surrounded by a substantial infalling circumstellar envelope \citep{ lada_star_formation_stages_1987, Hogerheijde_envelope_struc_1998}. The atomic jet is traced by [\ion{Fe}{II}], while a wider angle molecular ``shell'' outflow is traced by \ce{H2}. However, along the spine of the atomic jet, we find tentative evidence of collimated high-velocity, blue-shifted ($\gtrsim$100 km s$^{-1}$) \ce{H2} emission (see Figures \ref{fig:h2_velocity_bins},\ref{fig:figspectral_diffs}). The wider molecular ``shell'' outflow meanwhile has a speed of <70 km s$^{-1}$. These results suggest that co-spatial molecular and atomic jets are present in TMC1A, and, to the best of our knowledge, this is the first detection of a molecular jet in this source.

In TMC1A, the atomic jet remains significantly brighter and faster (by $\sim $50 km/s) than its molecular jet counterpart. This would suggest that TMC1A is in the tail end of the transition from a collimated jet that is composed mostly of molecular gas (Class 0 sources) to being mainly atomic (as in Class II sources) \citep[see, e.g.,][]{panoglouetal2012, Rabenanahary_wide_angle_outflows_driven_by_jet_2022}. This could be because the protostellar envelope and/or a large enough column of dust and gas in the outflow shielding the \ce{H2} jet from UV photo-dissociation or that the jet is fast enough to destroy most, but not all, of the \ce{H2} through shocks \citep{Tabone_H2_formation_shielding_2020}. To better understand such differences, deep maps of the atomic and molecular outflow in the near- and mid-IR for a large sample of protostars in different evolutionary stages are essential to better understand how the physical and chemical structure evolves with age.

\subsection{Intensity Variations in the [Fe II] and \ce{H2} Jets} 

Intensity variations and knot-like features have been seen in several protostellar outflows \citep[e.g.][]{hartiganetal2011_hhjets}. In our observations, we observe three distinct flux density peaks in both the blue-shifted atomic and molecular jets. Interestingly, we find that the [\ion{Fe}{II}] peaks are not quite co-spatial with the \ce{H2} peaks: The [\ion{Fe}{II}] line peaks at de-projected distances of 200, 270, and 370 AU while the \ce{H2} line peaks at de-projected distances of 245, 350, and 410 AU. It appears that the slower ($\sim$ 110 km s$^{-1}$ or 190 km s$^{-1}$ de-projected) \ce{H2} peaks are situated in between the faster ($\sim$ 150 km s$^{-1}$ or 260 km s$^{-1}$ de-projected) [\ion{Fe}{II}] peaks. While the intensity variations in TMC1A are not spatially well-resolved, they are similar to the knot-like features in some protostellar jets \citep[e.g.][]{ray2023outflows, toledano-juarezetal2023_l1448c} that are sometimes associated with Herbig-Haro (HH) objects, although historically at much larger ($\sim$ tenth of a parsec) scales \cite{reipurthbally01}.

The velocity differences between the \ce{H2} and [\ion{Fe}{II}] peaks indicate that these features could be produced by a faster atomic jet plowing into a slower molecular outflow. This could explain why the slower ($\sim$ 100 km s$^{-1}$) peaks in \ce{H2} are observed to be in between the faster moving ($\sim 150$ km s$^{-1}$) peaks in [\ion{Fe}{II}]. In this scenario, molecular material gets excited by shocks in the atomic [\ion{Fe}{II}] jet, which then cool, resulting in the \ce{H2} emission peaking in these regions.

The observed peaks may also be attributed to episodic mass ejection events \citep[e.g.][]{bonito2010} where the formation of chains of knots could be due to a variable mass-ejection rate from the protostar and disk \cite[e.g.][]{arce2007}, which is closely linked to episodic accretion onto the protostar \citep[e.g.][]{plunkett2015nature, vorobyov2018}. The intensity variations observed in TMC1A resemble the knots observed in other ``microjets" such as in DG Tau, where knots are inferred to form on timescales of just a few years \citep{aamboage11} or 10-15 years in the case of TH 28 \citep{murphy_TH28_microjet_2021}. In the episodic mass ejection scenario, the \ce{H2} emission peaks appearing in the valleys of the [\ion{Fe}{II}] emission in TMC1A would correspond to periods in between ejection events. In either case, higher spatial and spectral resolution observations, and/or several epochs of observations, will be necessary to confirm such a scenario.

Since the dynamical timescales associated with each successive observed peak along the jet axis are relatively small ($< 10$ years), periodically observing the outflow could reveal not only the movement of the detected intensity peaks, but also the formation of any new peaks in both [\ion{Fe}{II}] and \ce{H2}. This would inform us about the interaction and evolution of the atomic and molecular knot-like structures.  With {\it JWST} in particular, the simultaneous monitoring of near-IR atomic H recombination lines (such as H$\beta$ at 1.282 $\mu$m and H$\alpha$ at 4.052 $\mu$m) would reveal changes in the protostellar accretion rate and its relation to the evolution and possible creation of new "knot"-like structures. This would further our understanding of the dynamic processes driving the redistribution of material in protostellar systems as a function of time.

\section{Conclusions}\label{sec:conclusion} 

We analyze near-IR $JWST$ observations of [\ion{Fe}{II}] and \ce{H2} lines detected toward the Class I protostar, TMC1A, to study the structure of its protostellar outflows. The spectra of the TMC1A outflow show numerous [\ion{Fe}{II}] \& \ce{H2} lines. In total, we detect 30 [\ion{Fe}{II}] lines and 26 \ce{H2} lines in our NIRSpec IFU observations with the G140H, G235H and G395H gratings (see Figures \ref{fig:full_iron2lines_spectra}-- \ref{fig:full_H2lines_moment0}). Our findings can be summarized as follows:
\begin{itemize}
    \item The observed [\ion{Fe}{II}] 1.644 $\mu$m emission peaks at around -70 km s$^{-1}$ and reach velocities upwards of -150 km s$^{-1}$, corresponding to approximately -120 and -260 km s$^{-1}$ after de-projection. For the first time, the red-shifted counterpart of the atomic jet is seen, which is only detected in the bright 1.644 and 1.810 $\mu m$ [\ion{Fe}{II}] lines. The southern red-shifted emission is significantly dimmer (by a factor 10-30x) and faster than the blue-shifted component (peaking at $\sim$190 km s$^{-1}$ with tentative emission at 230 km s$^{-1}$ or $\sim$330 and $\sim$400 km s$^{-1}$ after de-projection). The atomic jet in TMC1A is bipolar.

    \item We find tentative evidence of a fast collimated \ce{H2} outflow with emission above 3-$\sigma$ at radial velocities of $\sim$110 km s$^{-1}$ ($\sim$190 km s$^{-1}$ deprojected) on the blue-shifted side of the outflow. It indicates that the jet toward TMC1A is also molecular in addition to the wider angle (slower) \ce{H2} shell. 
    
    \item Similar to [\ion{Fe}{II}] emission, \ce{H2} emission is also detected in the southern part of the protostar indicating that it is also bipolar and tends to be more red-shifted than the northern outflow. The \ce{H2} outflow is 2-4$\times$ wider in both the northern and southern outflows than the atomic [\ion{Fe}{II}] jet and the southern side is $\gtrsim2 \times$ wider than the northern counterpart at the same deprojected distance.
    
    \item Both [\ion{Fe}{II}] and \ce{H2} high-velocity emission ($v\sim 100-150$ km/s) show three distinct peaks that are not co-spatial. From the deprojected velocities and distances, [\ion{Fe}{II}] emissions peak at 200, 270, and 370 AU. On the other hand, \ce{H2} emission peaks at 245, 350, 410 AU. Furthermore, at the peak locations, \ce{H2} is observed to be slower than the closest [\ion{Fe}{II}] emission. 

    \item With the detected [\ion{Fe}{II}] lines, we derive a range of visual extinction from $A_{\rm V}=10-30$ that decreases with distance from the central protostar for the northern blue-shifted part. For the red-shifted jet, a comparable extinction is derived from 1.257/1.644 $\mu$m line ratio pair.


    \item The large difference in [\ion{Fe}{II}] line flux and velocity between the red-shifted and blue-shifted jet components may indicate that the red-blue asymmetry commonly observed in more evolved Class II jets can already start in the earlier Class I phase. 
    We showed the blue-red brightness asymmetry is not due to differential extinction, but is rather an intrinsic brightness difference.
    
    \item The collimated atomic and molecular jet in TMC1A illustrates a scenario in which molecular jets, typical in Class 0 stages, and atomic jets, prevalent in Class II stages where molecular outflows are observed at wider angles, coexist in the intermediate Class I stage. The presence of a brighter and faster atomic jet compared to the molecular jet, including the slower, wider-angle molecular shell, suggests that TMC1A may be near the tail end of this transitional phase of outflows.

    \item The peaks in flux profiles within the atomic and molecular jet resemble Herbig-Haro object knots but at smaller scales (a few hundred AU vs. parsecs), with the slower molecular peaks and offset disparities suggesting the possibility of the atomic jet driving the molecular jet. Their dynamical timescales, less than 10 years and with only a few years' difference, imply recent or ongoing outbursts.
\end{itemize}

The observations of TMC1A presented here are a great example of the NIRSpec IFU's ability to detect a wealth of [\ion{Fe}{II}] and \ce{H2} lines in the near-IR, even in the case of the highly extincted environments of protostars and, in the case of TMC1A, revealing high-velocity outflow components and brightness asymmetries. In future work, we will examine the range of excitation conditions that can be constrained by the large number of forbidden iron lines and \ce{H2} lines detected here, which will permit us to better characterize the physical conditions of the outflow and its respective mass outflow rates.

\begin{acknowledgements}
This work was supported in part by an ALMA SOS award, STScI JWST-GO-02104.002-A, NSF grant AST-1910106 and NASA ATP grant 80NSSC20K0533.  This work is based on observations made with the NASA/ESA/CSA James Webb Space Telescope.  The data were obtained from the Mikulski Archive for Space Telescopes at the Space Telescope Science Institute, which is operated by the Association of Universities for Research in Astronomy, Inc., under NASA contract NAS 5-03127 for JWST. These observations are associated with GO program \#2104.
DH is supported by a Center for Informatics and Computation in Astronomy (CICA) grant and grant number 110J0353I9 from the Ministry of Education of Taiwan. DH also acknowledges support from the National Science and Technology Council of Taiwan through grant number 111B3005191. LT acknowledges support from the Netherlands Research School for Astronomy (NOVA). A portion of this research was carried out at the Jet Propulsion Laboratory, California Institute of Technology, under a contract with the National Aeronautics and Space Administration (80NM0018D0004). We thank the referee for a constructive and timely report that helped improve the paper.
\end{acknowledgements}

%
%
\bibliographystyle{aa}
\bibliography{ref}
\begin{appendix}
\onecolumn
\section{Detected Lines}
\label{App:detected lines}
The list of detected lines for both [\ion{Fe}{II}] and H$_2$ are reported in Table \ref{tab:combined_detections} with observational details and the properties of each line taken from the \href{https://linelist.pa.uky.edu/newpage/}{Atomic Line List version:3.00b5} and \href{https://physics.nist.gov/PhysRefData/ASD/lines_form.html}{NIST Atomic Spectra Database} for [\ion{Fe}{II}] and the \href{https://www.gemini.edu/observing/resources/near-ir-resources/spectroscopy/important-h2-lines}{Gemini Important H2 lines list} for \ce{H2} . We report the intensity line ratios with respect to prominent lines (1.644 for [\ion{Fe}{II}] and 2.122 for \ce{H2})

\begin{table*}
    \centering
    \caption{[Fe II] and H$_2$ Detections}
    \begin{tabular}{cc}
        \begin{minipage}[t]{0.5\textwidth}
            \setlength{\tabcolsep}{3pt}

            \centering
            \begin{tabular}{cccccc}
                \hline
               $\lambda $($\mu m$)&S$_1$-S$_2$&A$\times 10^{-3}$ (s$^{-1}$)& $E_{upper}$ (K)&$F_{line}/F_{1.644}$   \\
               \hline
                \multicolumn{5}{c}{a2G-a4D} \\
                \hline
                1.0324$^1$, 1.0321$^2$& 7/2-5/2 & 2.5$^2$ & 26055$^{1,2}$& 0.003-0.008 \\
                
                \hline
                \multicolumn{5}{c}{a4D-a2G} \\
                \hline
                1.1885$^1$& 7/2-7/2 & -- & 16369$^1$ & 0.005-0.006\\
                \hline
                \multicolumn{5}{c}{a6D-a4D} \\
                \hline
                1.2489$^1$,1.2485$^{2}$& 7/2-5/2 & 0.4$^2$ &  8392$^{1,2}$ & 0.002-0.005 \\
                1.2570$^1$,1.2567$^2$&9/2-7/2&4.7$^1$,5.5$^2$&11445$^1$,7955$^2$&0.062-0.216\\
                1.2707$^1$,1.2703$^2$&1/2-1/2&3.3$^1$,3.6$^2$&8847$^{1,2}$&0.012-0.016\\
                1.2791$^1$,1.2788$^2$&3/2-3/2&2.5$^1$,3.6$^2$&12489$^1$,8680$^2$&0.016-0.021\\
                1.2946$^1$,1.2943$^2$&5/2-5/2&2.0$^1$,2.1$^2$&12074$^1$,8392$^2$&0.024-0.042\\
                1.2981$^1$,1.2978$^2$&1/2-3/2&1.1$^1$,1.2$^2$&12489$^1$,8680$^2$&0.007-0.012\\ 
                1.3209$^1$,1.3206$^2$&7/2-7/2&1.3$^1$,1.8&11445$^1$,7955$^2$&0.031-0.093\\
                1.3281$^1$,1.3278$^2$&3/2-5/2&1.2$^1$,1.5$^2$&12074$^1$,8392$^2$&0.014-0.030\\ 
                1.3722$^1$,1.3718$^2$&5/2-7/2&0.94$^2$&11445$^1$,7955$^2$&0.032-0.066
                \\ 
                \hline
                \multicolumn{5}{c}{a4F-a4D} \\
                \hline
                1.5339$^1$,1.5335$^2$&9/2-5/2&3.1$^1$,2.1$^2$&12074$^1$,8392$^2$&0.107-0.198\\ 
                1.5999$^1$,1.5995$^2$&7/2-3/2&4.2$^1$,4.4$^2$&12489$^1$,8680$^2$&0.096-0.226\\ 
                1.6440$^1$,1.6434$^2$&9/2-7/2&6.0$^1$,4.6$^2$&11445$^1$,7955$^2$&1.0\\ 
                1.6642$^1$,1.6638$^2$&5/2-1/2&4.7$^1$,4.0$^2$&12728$^1$,8847$^2$&0.076-0.179\\
                1.6773$^1$,1.6769$^2$&7/2-5/2&2.5$^1$,1.8$^2$&12074$^1$,8392$^2$&0.123-0.287\\ 
                1.7116$^1$,1.7111$^2$&5/2-3/2&1.2$^1$,2.0$^2$&12489$^1$,8680$^2$&0.048-0.097\\
                1.7454$^1$,1.7449$^2$&3/2-1/2&2.5$^1$,2.1$^2$&12728$^1$,8847$^2$&0.062-0.18\\
                1.7976$^1$,1.7971$^2$&3/2-3/2&2.1$^1$,1.1$^2$&12489$^1$,8681$^2$&0.077-0.258 \\ 
                1.8005$^1$,1.8000$^2$&5/2-5/2&1.8$^1$,1.6$^2$&12074$^1$,8392$^2$&0.148-0.461\\ 
                1.8099$^1$,1.8094$^2$&7/2-7/2&1.3$^1$,1.0$^2$&11445$^1$,7955$^2$&0.317-0.560\\ 
                1.8959$^1$,1.8954$^2$&7/2-5/2& 0.25$^2$ & 8391$^{1,2}$&0.029-0.062\\
                1.9541$^1$, 1.9536$^2$&5/2-7/2& 0.12$^2$ & 7955$^{1,2}$&0.043-0.056\\
                \hline
                \multicolumn{5}{c}{a4D-a4P} \\
                \hline
                1.7489$^1$,1.7484$^2$&7/2-3/2&1.9$^2$&13673$^{1,2}$&0.012-0.057\\ 
                1.8119$^1$,1.8114$^2$&7/2-5/2&2.2$^2$&13474$^{1,2}$&0.017-0.057\\ 
                \hline
                \multicolumn{5}{c}{a4P-a2P} \\
                \hline
                2.0073$^{1,2}$&1/2-1/2&102$^2$&18886$^{1,2}$&0.023-0.397\\
                2.0466$^{1,2}$&5/2-3/2&80.4$^2$&18361$^{1,2}$&0.022-0.058\\
                2.1334$^{1,2}$&3/2-3/2&39$^2$&18360$^{1,2}$&0.012-0.080\\  
                \hline
                \multicolumn{5}{c}{a2G-a2H}\\
                \hline
                2.0157$^{1,2}$&9/2-9/2&56$^2$&20805$^{1,2}$&0.015-0.296\\
                
                \hline
                \multicolumn{5}{c}{a6D-a4F} \\
                \hline
                4.8891$^1$,4.8891$^2$ &7/2-7/2&0.09$^2$&2430$^{1,2}$&1.16-2.31\\
            \end{tabular}
        \end{minipage}
        &
        \begin{minipage}[t]{0.48\textwidth}
        \setlength{\tabcolsep}{2pt}
            \centering
            \begin{tabular}{ccccc}
                \hline
                 $\lambda$ ($\mu m$)&Level&A$\times 10^{-7}$ (s$^{-1}$)&$E_{\text{upper}}$ (K)&$F_{line}/F_{2.122}$\\
                \hline
                \multicolumn{5}{c}{1-0 S}\\
                \hline
                1.7480&(7)&2.98&12817& 0.051-0.124 \\ 
                1.7880&(6)&3.54&11522& 0.048-0.075 \\ 
                1.8358&(5)&3.96&10341&0.262-0.469\\ 
                1.8920&(4)&4.19&9286&0.077-0.183\\ 
                1.9576&(3)&4.21&8365&0.469-0.778\\
                2.0338&(2)&3.98&7584&0.159-0.364\\
                2.1218&(1)&3.47&6956&1.0\\ 
                2.2235&(0)&2.53&6471&0.252-0.460\\ 
                \hline
                \multicolumn{5}{c}{2-1 S}\\
                \hline
                2.0735&(3)&5.77&13890&0.034-0.136\\
                2.2247&(1)&4.98&12550&0.046-0.158\\ 
                \hline
                \multicolumn{5}{c}{1-0 Q} \\
                \hline
                2.4237&red{(3)}&2.78&6956&0.843-1.58\\
                2.4375&(4)&2.65&7586&0.240-0.567\\
                2.4548&(5)&2.55&8365&0.828-2.58\\ 
                2.4756&(6)&2.45&9286&0.170-0.318\\ 
                2.5001&(7)&2.34&10341&0.342-0.529\\ 
                \hline
                \multicolumn{5}{c}{1-0 O}\\
                \hline
                2.6269&(2)&8.54&5987&0.570-1.55\\ 
                2.8025&(3)&4.23&6149 &1.73-3.08\\ 
                3.0039&(4)&2.90&6471&0.236-0.414\\ 
                3.2350&(5)&2.09&6956&0.466-1.02\\ 
                3.5007&(6)&1.50&7485 &0.128-0.377\\ 
                3.8075&(7)&1.06&8365&0.252-0.652\\ 
                \hline
                \multicolumn{5}{c}{0-0 S}\\
                \hline
                3.8464&(13)&16.2&17458&0.383-1.06\\
                4.1810&(11)&9.64&13703& 0.905-1.44\\
                4.4096&(10)&7.03&11940&0.695-1.20\\
                4.6947&(9)&4.90&10263&2.20-3.60\\
                5.0529&(8)&3.24&8677&1.15-3.12\\ 
                
                \hline
            \end{tabular}
            
        \end{minipage}
    \end{tabular}
    \tablefoot{\textbf{Left:} Forbidden [Fe II] lines detected in our observations with the vacuum/rest wavelength ($\lambda$), electronic spin states (S$_1$-S$_2$) in each respective line transitions (separated by horizontal lines with name of each level's electronic configuration), the Einstein A-coefficients (A), the lower and upper energy level (E$_{range}$), the range of intensity line ratios taken with respect to the 1.644 $\mu$m line using peak line intensity ($F_{line}/F_{1.644}$). We use both the \href{https://linelist.pa.uky.edu/newpage/}{Atomic Line List version:3.00b5} updated on 10/02/2023 and \href{https://physics.nist.gov/PhysRefData/ASD/lines_form.html}{NIST Atomic Spectra Database} for the Einstein A coefficients and upper energy. Both databases report slightly different values for rest wavelengths, Einstein A-coeffs, and upper energy levels. In many cases, only NIST reports the Einstein A. Values with upper script "1" are from the atomic line list and "2" are from NIST. For the case of the 1.1885 $\mu$m line, we note that only the atomic line list reports this line and no Einstein A-coefficient is reported. It is possible this is a different line such as [P II] as discussed in \cite{olivia_iron_phosphorus_2001}. \textbf{Right:} H$_2$ lines detected in our observations with the vacuum/rest wavelength ($\lambda$), rotational level (Level) in each respective ro-vibrational transition (separated by horizontal lines with name of the ro-vibrational transition), and the Einstein A-coefficient (A) derived in \citet{turner1977quadrupole},the upper energy level ($E_{\text{upper}}$) taken from \citet{dabrowski1984lyman}, the range of intensity line ratios taken with respect to the 2.122 $\mu$m line using peak line intensity ($F_{line}/F_{2.122}$). The properties of each line besides the last column are taken from the \href{https://www.gemini.edu/observing/resources/near-ir-resources/spectroscopy/important-h2-lines}{Gemini Important H2 lines list}.
    }
    \label{tab:combined_detections}
\end{table*}

\end{appendix}

\end{document}